%% file: dkstarlnu.tex
\documentclass[floatfix,showpacs,showkeys,aps,prl,twocolumn]{revtex4-2}
\usepackage{lineno}
\usepackage{amsmath}
\usepackage{amssymb}

\usepackage{tikz}
\usepackage{lipsum,adjustbox}
\usepackage{graphicx}
\usepackage[caption = false]{subfig}
\usepackage{dcolumn}
\usepackage{bm}
\usepackage{gensymb}

\usepackage{tabstackengine}
\usepackage[inline]{enumitem}

\setstackEOL{\cr}
\setstackgap{L}{\normalbaselineskip}

\let\oldequation\equation
\let\oldendequation\endequation

\renewenvironment{equation}
  {\linenomathNonumbers\oldequation}
  {\oldendequation\endlinenomath}

\newcolumntype{L}[1]{>{\raggedright\arraybackslash}p{#1}}
\newcolumntype{C}[1]{>{\centering\arraybackslash}p{#1}}
\newcolumntype{R}[1]{>{\raggedleft\arraybackslash}p{#1}}

\usepackage{hyperref}
\hypersetup{
 pdftitle={},
 pdfauthor={},
 pdfsubject={},
 pdfkeywords={},
 pdfstartview={},
 bookmarksopen=true, breaklinks=true, debug=true,
 colorlinks=true, linkcolor=blue, citecolor=red, urlcolor=blue,
 hyperfigures=true
}
\usepackage{todonotes}

\begin{document}

\title{\boldmath Observation of $D^+\to K^0_S\pi^0\mu^+\nu_\mu$, Test of Lepton Flavor Universality and First Angular Analysis of $D^+\to \bar{K}^\ast(892)^0\ell^+\nu_\ell$}

\input{authorlist_2025-01-22}

\begin{abstract}
	We report a study of the semileptonic decays $D^+\to K_S^0\pi^0\ell^+\nu_\ell$ ($\ell = e, \mu$) based on $20.3\,\mathrm{fb}^{-1}$ of $e^+e^-$ collision data collected at the center-of-mass energy of 3.773 GeV with the BESIII detector.
    The $D^+\to K_S^0\pi^0\mu^+\nu_\mu$ decay is observed for the first time, with a branching fraction of $(0.896\pm0.017_{\rm stat}\pm0.008_{\rm syst})\%$, and the branching fraction of $D^+\to K_S^0\pi^0e^+\nu_e$ is determined with the improved precision as $(0.943\pm0.012_{\rm stat}\pm0.010_{\rm syst})\%$.
    From the analysis of the dynamics, we observe that the dominant $\bar{K}^\ast(892)^0$ component is accompanied by an $S$-wave contribution, which accounts for $(7.10 \pm 0.68_{\rm stat} \pm 0.41_{\rm syst})\%$ of the total decay rate of the $\mu^+$ channel and $(6.39 \pm 0.17_{\rm stat} \pm 0.14_{\rm syst})\%$ of the $e^+$ channel. Assuming a single-pole dominance parametrization, the hadronic form factor ratios are extracted to be $r_V=V(0)/A_1(0)=1.42 \pm\, 0.03_{\rm stat} \pm\, 0.02_{\rm syst}$ and $r_2=A_2(0)/A_1(0)=0.75 \pm\, 0.03_{\rm stat} \pm\, 0.01_{\rm syst}$.
    Based on the first comprehensive angular and the decay-rate $CP$ asymmetry analysis, the full set of averaged angular and $CP$ asymmetry observables are measured as a function of the momentum-transfer squared; they are consistent with expectations from the standard model. No evidence for violation of $\mu-e$ lepton-flavor universality is observed in either the full range or the five chosen bins of momentum-transfer squared.
\end{abstract}

\maketitle

Experimental studies of semileptonic decays of heavy mesons provide important opportunities for rigorously testing the standard model (SM) and probing for potential new physics (NP). In recent years, some anomalies have been observed in semileptonic $B$ decays mediated by the charged current $b \to c \ell\bar{\nu}_{\ell}$, deviating from SM predictions by 2 to 3 deviations~\cite{Bphysicreview1,Bphysicreview2}, indicating possible violations of lepton flavor universality (LFU). Various NP scenarios, including $W^{\prime}$ models~\cite{wp1,wp2}, leptoquark models~\cite{prd2013leptonquark}, and charged Higgs models~\cite{prd2018chargehisgs}, proposed to resolve these tensions---can modify both branching fractions (BFs) and detailed angular distributions. Key observables therefore include BF ratios, and the full set of angular coefficients, which together constrain the relevant parameters.

By contrast, no anomalies have been observed experimentally in $D$ mesons to date~\cite{Zhangsfslide}, yet the same NP operators invoked to explain the $B$ anomalies~\cite{Belle2023bdstarlnu,Belle2024bdstarlnu} may impact the semileptonic $D$ decays~\cite{DNP1,DNP2,DNP3,DNP4,DNP5,DNP6}. Some observables---such as LFU violation~\cite{DNPShow1}, forward-backward asymmetry~\cite{DNPShow2} and lepton (hadron) polarization~\cite{DNPShow3}---have been proposed as sensitive probes. In particular, experimental analysis of the BF ratios and the angular distributions in $D^+\to \bar{K}^{*0} \ell^+\nu_\ell$~($\ell=e,\mu$) could reveal subtle NP effects~\cite{illustrate}.

The unique mass range of charmed hadronic states allows for rich nonperturbative hadronic physics~\cite{FPinQuark,Cabibbo:1963yz,Kobayashi:1973fv}, making precision studies of heavy-to-light form factors (FFs) a key component in testing the SM. The FFs are inherently nonperturbative and describe strong interaction effects in the initial and final hadrons; theoretical predictions within the SM can only be calculated by nonperturbative approaches, such as lattice quantum chromodynamics (LQCD)~\cite{FFLQCD1,FFLQCD2,FFLQCD3,FFLQCD4,FFLQCD5}, quark models~\cite{FFQuark1,FFQuark2,FFQuark3} and QCD sum rules~\cite{FFSumrule}. Precise experimental determination of these hadronic transition FFs and their comparison with theoretical predictions not only provides stringent tests of the SM, but also offers a means to constrain or validate NP models.

Experimental measurements of the BFs and hadronic transition FFs of $D^{+}\to \bar{K}^{*0}\ell^+\nu_\ell$~($\ell=e,\mu$) via $\bar{K}^{*0}\to K^-\pi^+$ have been reported by several experiments~\cite{E691,E653,E687,E791I,E791II,MARKIII1991,BEATRICE1999,CLEO2010,FOCUS2002,BaBar2010,BESanfenfen,Ke:2023qzc}. This Letter presents the first measurement of averaged angular observables and $CP$ asymmetries in the full set of angular distributions and asymmetries in the decay rates of $D^+\to \bar{K}^{*0}\ell^+\nu_\ell$ as well as the BF and FF measurements, via $\bar{K}^{*0}\to K_S^0\pi^0$. This $\bar K^{*0}$ decay mode avoids large misidentification rate between $\pi^+$ and $\mu^+$ for $\bar{K}^{*0}\to K^-\pi^+$. In addition, we test $\mu-e$ LFU for $D^+ \to \bar{K}^{*0} \ell^+ \nu_\ell$ in the full $q^2$ range and in five $q^2$ intervals.

Our measurement is based on five independent kinematic variables: the squared invariant masses of the dihadron and dilepton systems, $m^2\equiv m^2(K_S^0\pi^0)$ and $q^2\equiv m^2(\ell^+\nu_\ell)$, and three decay angles $\theta_\ell$, $\theta_K$, and $\chi$ (see Fig.~1 in Supplemental Material~\cite{supplemental}). The decay angles are defined as follows: $\theta_K$ is the angle between the momentum of the kaon in the $K \pi$ rest frame and the direction of the $K \pi$ system in the $D$ rest frame; $\theta_\ell$ is the angle between the momentum of the charged lepton in the $\ell \nu_\ell$ rest frame and the direction of the $\ell \nu_\ell$ system in the $D$ rest frame; and $\chi$ is the angle between the normals to the planes defined in the $D$ rest frame by the $K \pi$ and $\ell \nu_\ell$ pairs, respectively~\cite{ke4pipiphase}. When analyzing $D^-$ decays, the sign of $\chi$ is reversed.

The five-dimensional differential decay rate for $D^+\to K_S^0\pi^0\ell^+\nu_\ell$ can be expressed in terms of nine angular coefficients $\mathcal{I}_{1-9}$, which depend on $m^2, q^2$ and $\cos \theta_K$~\cite{ke4pipiphase,kl4pipiphase,bl4-dl4,Richman:1995wm,Dl4mass,Zhang:2023nnn}, as

\begin{equation}\small\label{eq:decay_rate}
    \begin{aligned}
    &\frac{d^5 \Gamma}{d m^2 d q^2 d \cos \theta_K d\cos\theta_\ell d \chi}= \frac{G_F^2|V_{c s}|^2}{(4 \pi)^6 m_D^3} \lambda \beta_m\beta_\ell\bigg(\mathcal{I}_{1}+\\
    &\mathcal{I}_{2}\cos2\theta_\ell+\mathcal{I}_{3}\sin^{2}\theta_\ell\cos2\chi+\mathcal{I}_{4}
    \sin2\theta_\ell\cos\chi+\mathcal{I}_{5}\sin\theta_\ell\cos\chi  \\
    &+\mathcal{I}_{6}\cos\theta_\ell+\mathcal{I}_{7}\sin\theta_\ell\sin\chi+\mathcal{I}_{8}\sin2\theta_\ell\sin\chi +\mathcal{I}_{9}\sin^{2}\theta_\ell\sin2\chi\bigg).
\end{aligned}
\end{equation}

The prefactor contains Fermi coupling constant $G_F$, CKM matrix element $V_{cs}$, and the mass of $D$ ($m_D$). Other parameters include $\lambda = p_{K\pi}m_{D}$, with $p_{K\pi}$ the $K\pi$ momentum in the $D$ rest frame, $\beta_m = 2p^*/m$, where $p^*$ is the kaon momentum in the $K\pi$ rest frame and $m$ the $K\pi$ invariant mass, and $\beta_\ell$ defined analogously to $\beta_m$ but for the $\ell\nu_\ell$ system.

Hadronic transition FFs are measured by performing an unbinned maximum likelihood fit to the decay rate expressed in terms of the SM helicity amplitudes~[see Eqs.~(6) and (7) in Supplemental Material~\cite{supplemental}]. The amplitude analysis includes components for the $\bar{K}^{*0}$ resonance and a $(K_S^0\pi^0)_{S-\text{wave}}$~[see Eqs.~(11) and (15) in Supplemental Material~\cite{supplemental}].

The asymmetry of the total decay widths, $\Gamma$, between $D^+$ and $D^-$ decays is defined as $A_{CP} \equiv \frac{\Gamma(D^+ \to \bar{K}^{*0} \ell^+\nu_\ell)-\Gamma(D^- \to K^{*0} \ell^-\bar{\nu}_\ell)}{\Gamma(D^+ \to \bar{K}^{*0} \ell^+\nu_\ell)+\Gamma(D^- \to K^{*0} \ell^-\bar{\nu}_\ell)}.$ The eight coefficients $\mathcal{I}_{2-9}$ can be obtained by defining piecewise angular-integration ranges in $\chi$ and $\cos \theta_\ell$. For example, the unnormalized coefficient $\mathcal{I}_{2}$ is obtained as
\begin{equation}\label{eq:integral}
    \begin{aligned}
        &\mathcal{I}_2= \int_{-\pi}^\pi d\chi\left[\int_{-1}^{-0.5} d\cos\theta_\ell+\int_{0.5}^1 d\cos\theta_\ell\right.\\
        &\left.\hspace{5.5em}-\int_{-0.5}^{0.5} d\cos\theta_\ell\right] \frac{d^5 \Gamma}{d q^2 \, dm^2 \, d \cos \theta_K d\cos\theta_\ell d \chi}.\\
    \end{aligned}
\end{equation}
Corresponding integration ranges to obtain unnormalized $\mathcal{I}_{3-9}$ are listed in Supplemental Material~\cite{supplemental}. Since the term $\mathcal{I}_1$ has no dependence on the decay angles, it provides only a normalization factor and is not considered in this Letter. The normalized and integrated observables $\langle \mathcal{I}_i\rangle$ are defined as
\begin{equation}\scriptsize\footnotesize\label{eq:integralcosK}
    \begin{aligned}
    &\langle \mathcal{I}_{2,3,6,9}\rangle=\frac{1}{\Gamma} \int_{q_{\min }^2}^{q_{\max }^2} d q^2 \int_{m_{\min }^2}^{m_{\max }^2} dm^2 \int_{-1}^{+1} d \cos \theta_K \, \mathcal{I}_{2,3,6,9},\\
    &\langle \mathcal{I}_{4,5,7,8}\rangle=\frac{1}{\Gamma} \int_{q_{\min }^2}^{q_{\max }^2} d q^2 \int_{m_{\min }^2}^{m_{\max }^2} dm^2 \left[\,\int_0^{+1} d \cos \theta_K\right.\\
    &\left.\hspace{15em}-\int_{-1}^0 d \cos \theta_K\right] \mathcal{I}_{4,5,7,8}.\\
    \end{aligned}
\end{equation}

In addition to the full $q^2$ range, five different integration ranges (bins) in $q^2$ are defined. The integration in $\cos \theta_K$ is defined to optimize the sensitivity to beyond-SM effects by integrating out contributions from the dominant $P$-wave $\bar{K}^{*0}$ resonance in the dihadron system, the integration regions differ since $\mathcal{I}_{2,3,6,9}$ are even in $\cos\theta_K$, while $\mathcal{I}_{4,5,7,8}$ are odd. Experimentally, the observables are determined by measuring the decay-rate asymmetries in the data split according to piecewise-defined angular regions. As an example, from Eqs.~(\ref{eq:integral}) and (\ref{eq:integralcosK}), $\langle \mathcal{I}_2\rangle$ is measured as
\begin{equation}
    \left\langle \mathcal{I}_2 \right\rangle=\frac{1}{\Gamma} \left[\Gamma\left(\left|\cos \theta_\ell\right|>0.5\right)-\Gamma\left(\left|\cos \theta_\ell\right|<0.5\right)\right].
\end{equation}
The observables $\langle \mathcal{I}_i\rangle$, measured separately for $D^+$ and $D^-$ mesons, are labeled as $\langle \mathcal{I}_i\rangle$ and $\langle \bar{\mathcal{I}}_i\rangle$, respectively. Their $CP$ average $S_i$ and asymmetry $A_i$ are defined as $\langle S_i\rangle= \left[ \langle \mathcal{I}_i\rangle + \langle \bar{\mathcal{I}}_i\rangle \right]$ and $\langle A_i\rangle= \left[ \langle \mathcal{I}_i\rangle - \langle \bar{\mathcal{I}}_i\rangle \right]$.
The well-known forward-backward asymmetry $A_{\rm FB}$~\cite{Belle2006AFB} and triple-product asymmetry $A_{\rm TP}$~\cite{Gronau2011ATP} are related to $\langle S_6\rangle$ and $\langle S_9\rangle$, respectively. If only SM amplitudes contribute to the decay processes, the $CP$-averaged observables $\langle S_{7,8,9}\rangle$ are predicted to vanish and constitute SM null tests together with the $CP$ asymmetries $\langle A_{2-9} \rangle$.

The BESIII detector~\cite{BES3,Yu:2016cof} consists of a helium-based multilayer drift chamber~(MDC), a plastic scintillator time-of-flight system~(TOF), and a CsI(Tl) electromagnetic calorimeter~(EMC), which are all enclosed in a superconducting solenoidal magnet providing a 1.0~T magnetic field. The inclusive Monte Carlo (MC) sample, described in Refs.~\cite{geant4,kkmc,evtgen,lundcharm,photos}, detailed in Supplemental Material~\cite{supplemental}, has been validated for background simulation. The signal decays $D^+\to K_S^0\pi^0\ell^+\nu_\ell$ are generated with the helicity amplitude of $\bar{K}^{*0}$ and LASS formalism of $(K_S^0\pi^0)_{S-\text{wave}}$ \cite{Swavetheory,Swavelass}, with parameters obtained in this Letter.

At $\sqrt{s}=3.773$ GeV, the $D\bar{D}$ meson pairs are produced without accompanying particles in the final state. This property allows us to study semileptonic $D$ decays with the double-tag~(DT) method. In this method, a single-tag (ST) event is one in which a $D^-$ meson is fully reconstructed via one of the six hadronic decay modes $K^+\pi^-\pi^-,K_S^0\pi^-$, $K^+\pi^-\pi^-\pi^0$, $K_S^0\pi^-\pi^0$, $K_S^0\pi^+\pi^-\pi^-$, and $K^+K^-\pi^-$. A DT event is selected by reconstructing the signal-side decay $D^+\to K_S^0 \pi^0 \ell^+ \nu_\ell$, where $K_S^0\to \pi^+\pi^-$, in the presence of an ST candidate. The BF of the signal decay is determined by $\mathcal{B}_{\rm sig} = \frac{N_{\rm DT}}{N^{\rm tot}_{\rm ST} \, \bar{\epsilon}_{\rm sig}}$, where $N^{\rm tot}_{\rm ST}=\sum_{j}{N^j_{\rm ST}}$  is the total number of ST events summed over all tag modes, and $N_{\rm DT}$ is the total number of DT events; $\bar{\epsilon}_{\rm sig}=\sum_{j}\frac{N^j_{\rm ST}}{N^{\rm tot}_{\rm ST}}\frac{\epsilon^j_{\rm DT}}{\epsilon^j_{\rm ST}}$ is the averaged signal efficiency of selecting $D^+\to K_S^0 \pi^0 \ell^+ \nu_\ell$ in the presence of the ST $D^-$ meson, where $\epsilon^j_{\rm ST}$ and $\epsilon^j_{\rm DT}$ are the ST and DT efficiencies for the $j$ th ST mode, respectively. A correction factor is applied to account for data-MC discrepancies in the signal efficiencies, ~\cite{supplemental}.

The selection criteria of $\pi^\pm,K^\pm,K_S^0,\gamma$, and $\pi^0$ candidates are widely used in our previous works~\cite{BEStaoluyankk,BESpanxetap}, and details are provided in Supplemental Material~\cite{supplemental}. Two kinematic variables, the energy difference $\Delta E \equiv E_{D^-}-E_{\rm beam}$ and the beam-constrained mass $M_{\rm BC}\equiv \sqrt{E_{\rm beam}^2/c^4-|\vec{p}_{D^-}|^2/c^2}$, are used to distinguish the ST $D^-$ mesons from combinatorial backgrounds, where $E_{\rm beam}$ is the beam energy and $(E_{D^-},\vec{p}_{D^-})$ is the four-momentum of the ST $D^-$ in the $e^+e^-$ rest frame. The combination with the smallest $|\Delta E|$ is chosen if multiple combinations are in an event.

The ST candidates are required to satisfy mode-dependent requirements on $\Delta E$ corresponding to $\pm 3.5\sigma$ around the fitted peaks. The ST yields in data for each ST mode are extracted by fitting individual $M_{\rm BC}$ distributions. Candidates with $M_{\rm BC}\in [1.863,1.877]$~GeV/$c^2$ are retained for further analysis. The $\Delta E$ requirements, the ST yields in data, and the ST efficiencies for different ST modes are listed in Supplemental Material~\cite{supplemental}. Summing over all ST modes yields a total of $N^{\rm tot}_{\rm ST}=(10646.9\pm4.0)\times10^3$ ST candidates.

In the presence of the ST $D^-$, candidates for $D^+\to K_S^0 \pi^0 \ell^+ \nu_\ell$ are selected from the remaining tracks and showers. The selection criteria of $K_S^0$, and $\pi^0$ candidates are the same as ST selection. The $\ell^+$ candidates are selected by performing PID with the specific ionization energy loss in the MDC ${\rm d}E/{\rm d}x$, time of flight, EMC information, based on which likelihoods for positron, muon, pion, and kaon hypotheses $\mathcal{L}_{e,\mu,\pi,K}$ are calculated. The $e^+$ is identified with $\mathcal{L}_{e}>0.8(\mathcal{L}_{\rm K}+\mathcal{L}_{e}+\mathcal{L}_{\pi})$ and $\mathcal{L}(e)>0.001$, while the $\mu^+$ candidate must satisfy $\mathcal{L}(\mu)>\mathcal{L}(e)$, $\mathcal{L}(\mu)>\mathcal{L}(K)$, $\mathcal{L}(\mu)>\mathcal{L}(\pi)$ and $\mathcal{L}(\mu)>0.001$. To further distinguish $\ell^+$ candidates from hadronic backgrounds, the requirements $E_{\rm EMC}^{\mu}<0.3$ GeV and $E_{\rm EMC}^{e}/(p\cdot c)\in(0.4,\,1.4)$ are imposed on $\mu^+$ and $e^+$ candidates, respectively, where $E_{\rm EMC}$ is the energy deposited in the EMC and $p$ is the MDC momentum. Furthermore, the maximum energy of the extra photon, $E_\gamma^{\max}$, is required to be less than 0.2~GeV for the $\mu^+$ channel and 0.5 GeV for the $e^+$ channel.

Since the neutrino cannot be directly detected, its four-momentum $(E_{\rm miss},\vec{p}_{\rm miss})$ is obtained by calculating the missing energy and momentum, defined as $E_{\rm miss}\equiv E_{\rm beam}-\sum_{k}E_k$ and $\vec{p}_{\rm miss}\equiv \vec{p}_{D^-}-\sum_k\vec{p}_{k}$. Here, the index $k$ sums over the $K_S^0,\pi^0$ and $\ell^+$ of the signal candidate and $(E_k,\vec{p}_{k})$ is the four-momentum of the $k$ th particle. A kinematic variable $U_{\rm miss}\equiv E_{\rm miss} - |\vec{p}_{\rm miss}|c$ is defined to extract the signal yield. To further improve the resolution of the missing momentum, the $D^-$ momentum is constrained as $\vec{p}_{D^-}=-\hat{p}_{D^-}\sqrt{E^2_{\rm beam}/c^2-m^2_{D^-}c^2}$, where $\hat{p}_{D^-}$ is the unit vector in the momentum direction of the ST $D^-$ meson, and $m_{D^-}$ is the nominal mass of the $D^-$ meson~\cite{pdg}.

The potential background from $D^+\to K^0_S\pi^+\pi^0$ is suppressed by requiring the $K^0_S\pi^0\ell^+$ invariant mass to satisfy $M_{K_S^0\pi^0\ell^+} <$ 1.6 (1.8)~GeV/$c^2$ for the $\mu^+$ ($e^+$) channel.

\begin{figure}[htbp]\centering
    \includegraphics[width=0.49\textwidth]{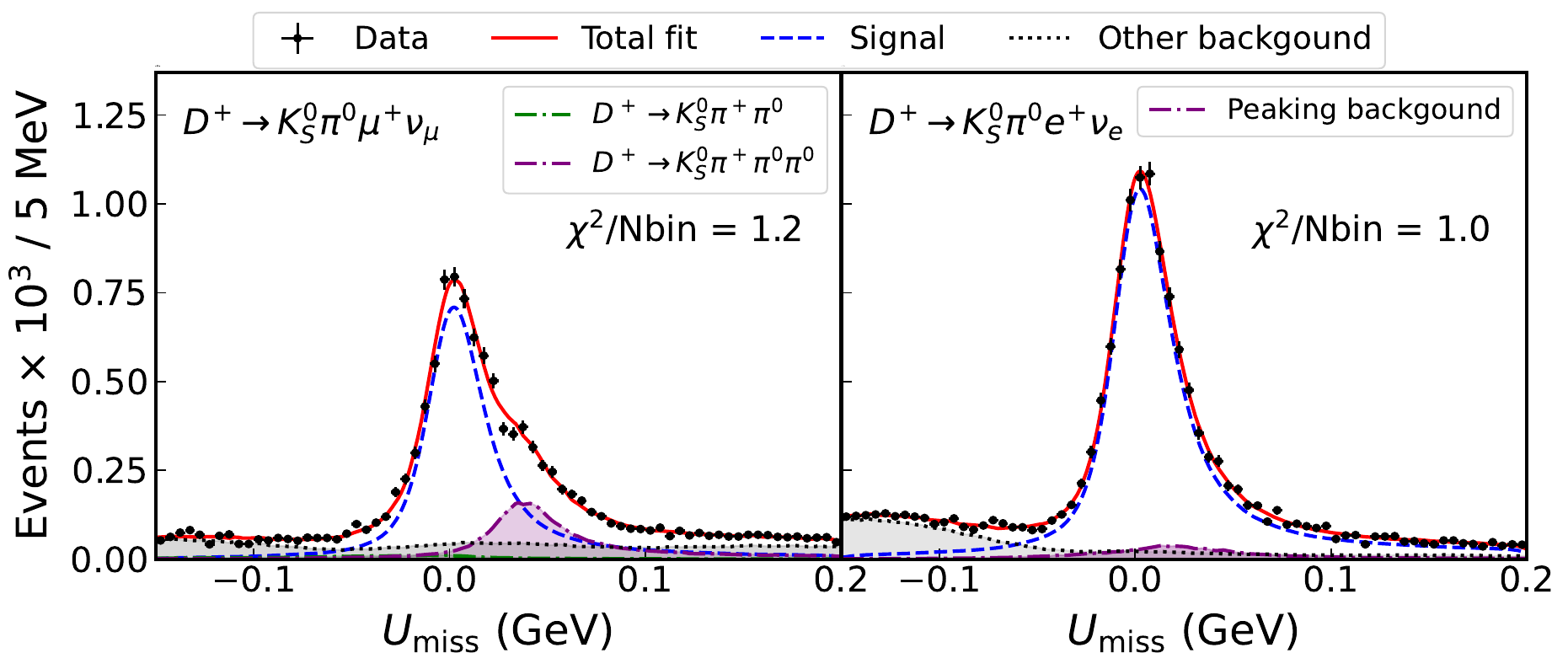}
    \caption{Fits to the $U_{\rm miss}$ distributions of the accepted candidates.}
    \label{fig:umissfit}
\end{figure}

The signal yields of $D^+\to K_S^0\pi^0 \mu^+\nu_\mu$ and $D^+\to K^0_S\pi^0e^+\nu_e$ are determined to be $6767\pm 126_\text{stat}$ and $11095\pm 137_\text{stat}$, from an unbinned maximum-likelihood fit on the $U_{\rm miss}$ distributions of the accepted candidates, as shown in Fig.~\ref{fig:umissfit}. In the fit, the signal shape is modeled using the MC simulation, convolved with a Gaussian resolution function with free parameters, and the background shape is derived from the inclusive MC sample. For the $\mu^+$ channel, the yield of peaking background $D^+\to K_S^0 \pi^+ \pi^0$ is fixed according to MC simulation, while the yields of both the $D^+\to K_S^0 \pi^+ \pi^0 \pi^0$ peaking background and other flatter backgrounds are separately floating. For the $e^+$ channel, the yields of peaking backgrounds $D^+\to K_S^0 \pi^+ \pi^0$, $D^+\to K_S^0 \pi^0\mu^+\nu_\mu$ and $D^0\to K_S^0 \pi^- e^+ \nu_e$ are fixed according to MC simulation, while the remaining background yield is floated. The first two are due to the misidentification of leptons, while the last one is due to particle exchange between tag and signal sides. The signal efficiencies are estimated to be $(10.44\pm 0.04_\text{stat})\%$ and $(17.09\pm 0.05_\text{stat})\%$ for the $\mu^+$ and $e^+$ channels, based on signal MC samples generated according to our amplitude analysis results for the helicity amplitudes of $\bar{K}^{*0}$ and the LASS formalism for the $(K_S^0\pi^0)_{S-\text{wave}}$. The BFs of $D^+\to K_S^0\pi^0 \mu^+\nu_\mu$ and $D^+\to K^0_S\pi^0e^+\nu_e$ are determined to be $0.896\pm 0.017_\text{stat}\pm 0.008_\text{syst}$ and $0.943\pm 0.012_\text{stat}\pm 0.010_\text{syst}$, respectively. 

The systematic uncertainties in the BF measurement are discussed below. The uncertainties of $\mu^+(e^+)$ tracking, 0.2\% (0.2\%), PID, 0.3\% (0.2\%), $K_S^0$ reconstruction, 0.2\% (0.2\%), and $\pi^0$ reconstruction, 0.2\% (0.2\%), are estimated using $e^+e^-\to \gamma\mu^+\mu^-$ events, $e^+e^-\to \gamma e^+ e^-$ events and DT $D^- D^+$ hadronic events, respectively. The uncertainties associated with the $E_\gamma^{\max}$, $N_{\rm extra}^{\rm track}$, and $N_{\rm extra}^{\pi^0}$ requirements and $M_{K_S^0\pi^0\ell^+}$ requirements are estimated to be 0.6\% (0.6\%) with a control sample of $D^+\to K_S^0\pi^+\pi^0(\pi^0)$. The uncertainty associated with the fit to the $U_{\rm miss}$ distribution is estimated to be 0.3\% (0.7\%) by varying the shapes and fixed normalizations. The common uncertainty in the ST $D^-$ yield is due to the fit to the $M_{\rm BC}$ distributions assigned as 0.3\% by varying the signal and background shapes. The uncertainty related to the signal MC model is taken as 0.2\% (0.2\%), which is the change of the DT efficiencies via varying the input parameters by $\pm 1\sigma$. Common uncertainties due to the limited MC statistics, 0.2\%, and the quoted BF of $K_S^0\to\pi^+\pi^-$, 0.1\%, are also included. The total systematic uncertainties are 0.9\% and 1.1\% for the $\mu^+$ and $e^+$ channels, respectively.

\begin{figure*}[htbp]\centering
    \includegraphics[width=0.98\textwidth]{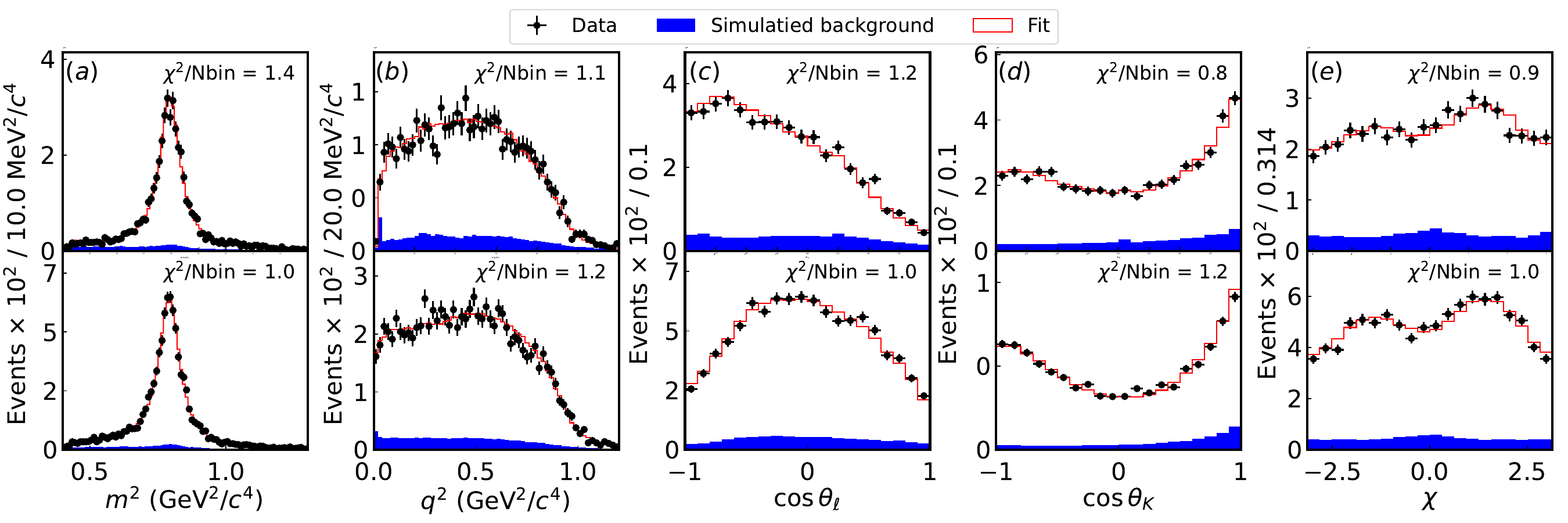}
    \caption{Projections of the data and amplitude analysis fits for the (top) $\mu^+$ and (bottom) $e^+$ channels.}
    \label{fig:pwaresult}
\end{figure*}

Using the SM helicity-amplitude expansion, we perform an unbinned maximum-likelihood fit to extract the hadronic transition FFs. To select high-purity samples, a strict requirement $|U_{\rm miss}|<0.02\,(0.05)$ GeV is used for the $\mu^+$ ($e^+$) channel. After selection, 4170 $\mu^+$ and 8929 $e^+$ candidate events survive with background fractions of $(13.3\pm 0.3)\%$ and $(9.1\pm 0.2)\%$, respectively. The helicity FFs are expressed in terms of two axial-vector FFs, $A_1(q^2)$ and $A_2(q^2)$, and one vector FF, $V(q^2)$. The FFs are all described with a simple pole form $A_{1,2}(q^2)=A_{1,2}(0)/(1-q^2/M^2_A)$ and $V(q^2)=V(0)/(1-q^2/M^2_V)$, with pole masses $M_V=1.81$~GeV/$c^2$ and $M_A=2.61$~GeV/$c^2$ fixed according to Ref.~\cite{BESanfenfen}. The FF ratios, $r_V=V(0)/A_1(0)$ and $r_2=A_2(0)/A_1(0)$, are determined from the fit. The $S$-wave amplitude parametrization is taken from Refs.~\cite{BESanfenfen,BaBar2010}, in which the scattering length, $a^{1/2}_{\rm S,BG}$, and the relative intensity, $r_S$, are determined by the fit. The dimensionless coefficient, $r_S^{(1)}$, and the effective range, $b^{1/2}_{\rm S,BG}$, are fixed to $0.08$ and $-0.81$~(GeV/$c$)$^{-1}$~\cite{BESanfenfen}. The background shape is modeled with the inclusive MC sample using the XGBoost algorithm~\cite{xgboost}. Simultaneous fits are performed to the $\mu^+$ and $e^+$ channels by minimizing the sum of their negative log-likelihood functions. Figure~\ref{fig:pwaresult} shows the projections of the nominal fit result. The fitted parameters and the fit fractions are summarized in Table~\ref{tab:ffparameters}.

\begin{table}[htbp]\centering
    \caption{The fitted parameters for the amplitude analysis, where the first and second uncertainties are statistical and systematic, respectively.}
    \begin{tabular}{lcc}
    \hline\hline
    Variable                                    & $D^+\to K_S^0\pi^0e^+\nu_e$  & $D^+\to K_S^0\pi^0\mu^+\nu_\mu$ \\
    \hline
    $m_{\bar{K}^{*0}}~(\text{MeV}/c^2)$         & \multicolumn{2}{c}{$897.3\pm0.3\pm2.2$}            \\
    $\Gamma_{\bar{K}^{*0}}~(\text{MeV}/c^2)$    & \multicolumn{2}{c}{$45.2\pm0.6\pm0.5$}             \\
    $r_{V}$                                     & \multicolumn{2}{c}{$1.42\pm0.03\pm0.02$}              \\
    $r_{2}$                                     & \multicolumn{2}{c}{$0.75\pm0.03\pm0.01$}              \\
    $a^{1/2}_{S,\text{BG}}~(\text{GeV}/c)^{-1}$ & $2.32\pm0.11\pm0.25$         & $1.47\pm0.25\pm0.22$   \\
    $r_{S}~(\rm GeV)^{-1} $                     & $-8.44\pm0.13\pm0.37$        & $-9.59\pm0.46\pm0.58$  \\
    $f_{S\text{-}{\rm wave}}$                   & $6.39\pm0.17\pm0.14$         & $7.10\pm0.68\pm0.41$   \\
    $f_{\bar{K}^{*0}}$                          & $93.50\pm0.18\pm0.28$        & $92.81\pm0.67\pm0.47$  \\
    \hline
    \hline
    \end{tabular}
    \label{tab:ffparameters}
\end{table}

The systematic uncertainties are evaluated as the difference between the nominal fit results and those obtained after changing a variable or a condition by an amount corresponding to the estimated uncertainty of that quantity. The detailed fitting systematic studies are discussed in Supplemental Material~\cite{supplemental}.

Each angular observable, and $A_{CP}$, is determined from the partial decay rate $d\Gamma_i/dq^2_i\,(\text{Tag})$, where ``Tag'' refers to the angular and flavor tags defined by the piecewise integration from Eqs.~(\ref{eq:integral}) and (\ref{eq:integralcosK}). The partial decay rate depends on the yield in each Tag and $q^2$ bin, and the yields are measured independently through the same fit methods used for the BF measurement but with the candidates splitted by Tag and $q^2$ region. The fits use the same modeling described earlier, including the same treatment of floating components, but with shapes determined independently in each Tag and $q^2$ region. The results for $\langle A_i \rangle$, $\langle S_i \rangle$, and $A_{CP}$, including both statistical and systematic uncertainties added in quadrature, are shown in Fig.~\ref{fig:asyAS}. A tabulated version is given in Supplemental Material~\cite{supplemental}. In general, the null-test observables show agreement with the SM predictions of zero.

\begin{figure*}[htbp]\centering
    \includegraphics[width=0.99\textwidth]{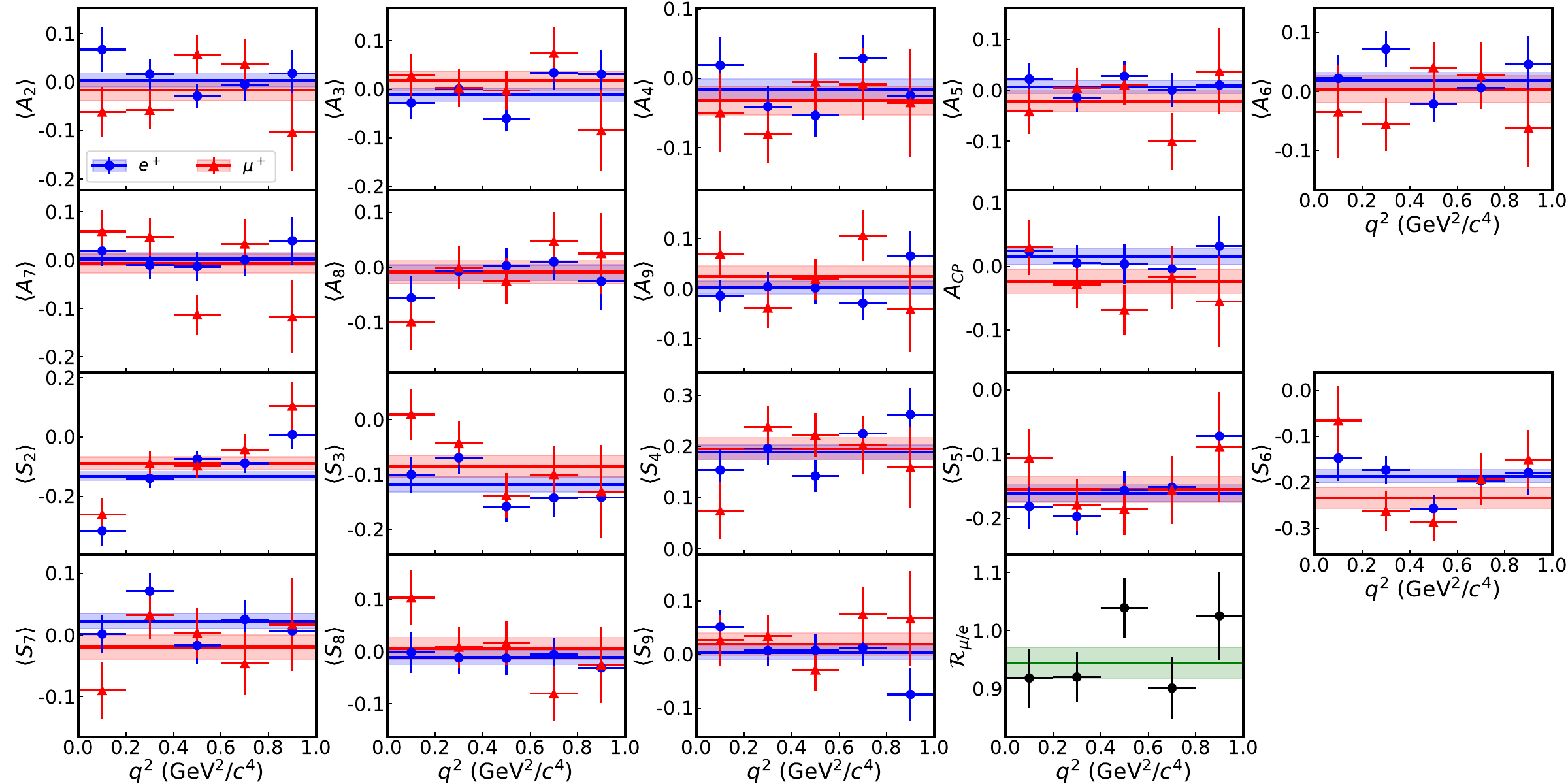}
    \caption{Measured observables $\langle A_i \rangle$, $\langle S_i \rangle$, $\mathcal{R}_{\mu/e}$ and the $CP$ asymmetry for $D^+ \to \bar{K}^{*0}\ell^+\nu_\ell$ in different $q^2$ regions with $0.8<m_{\bar{K}^{*0}}<1.0$ GeV$^2/c^4$. The $e^+$ ($\mu^+$) results are shown in blue (red). The horizontal lines and shaded bands represent the mean value and 1 standard deviation across full $q^2$ range.}
    \label{fig:asyAS}
\end{figure*}

The systematic uncertainties typically range from negligible to 50\% of the statistical uncertainty, depending on the observable and the $q^2$ region. These arise from the following sources: the model used in the $U_\text{miss}$ fits, the $D^-$ ST yield, the input $D^+$ lifetime, the estimation of the efficiency correction, and the limited size of the MC simulation sample. The leading systematic uncertainties are from the $U_\text{miss}$ fits; these are evaluated by repeating the analysis after either changing the simulation shape or fixed yield, the largest contribution from the background shape, or the fixed peaking background yield.

The ratio of the BFs between $D^+\to K_S^0\pi^0\mu^+\nu_\mu$ and $D^+\to K_S^0\pi^0e^+\nu_e$ is determined to be $\mathcal{R}_{\mu/e}=0.94 \pm 0.02_{\rm stat} \pm 0.01_{\rm syst}$, which is consistent with the SM predictions from 0.92 to 0.95~\cite{DNPShow3}. The values of $\mathcal{R}_{\mu/e}$ in different $q^2$ intervals are shown in Fig.~\ref{fig:asyAS}. Here, the correlated systematic uncertainties associated with the reconstruction of $K_S^0$ and $\pi^0$, $E_\gamma^{\max},\,N_{\rm extra}^{\rm track},\,N_{\rm extra}^{\pi^0}$ and $K_S^0\pi^0\ell^+$ requirements, as well as the quoted BF and MC model, largely cancel out. The results in $q^2$ bins are all consistent with the SM predictions.

In summary, using $20.3\,\mathrm{fb}^{-1}$~\cite{20lumi} of $e^+e^-$ collision data taken at $\sqrt{s}=$ 3.773 GeV with the BESIII detector, the first measurement of full set of $CP$ asymmetries and averaged angular observables with $A_{CP}$ in the decays $D^+\to \bar{K}^{*0} \ell^+\nu_\ell$ via $\bar{K}^{*0}\to K_S^0\pi^0$ is reported. All measured null-test observables $A_{CP}$, $\langle S_{7-9} \rangle$ and $\langle A_{2-9} \rangle$ are in agreement with the SM predictions, with the largest deviation being $1.7\sigma$. These results help constrain the parameters of physics models extending the SM. In addition, the absolute BF of $D^+\to K_S^0\pi^0\mu^+\nu_\mu$ is measured for the first time, and the absolute BF of $D^+\to K_S^0\pi^0e^+\nu_e$ is measured with improved precision. By analyzing the $D^+\to K_S^0\pi^0\ell^+\nu_\ell$ decay dynamics, the fraction of the $P\text{-wave}$ component and the hadronic FF ratios of $D^+\to \bar{K}^{*0}$ are presented. We have also tested LFU with $D^+ \to \bar{K}^{*0} \ell^+ \nu_\ell$ integrated over $q^2$ as well as in five $q^2$ intervals, and no LFU violation is observed.

\input{acknowledgement_2025-01-22}

\input{bibitem}
\onecolumngrid

\begin{center}
\section*{Supplemental Material}
\end{center}

\section{Definitions of the Decay Angles}
\begin{figure}[htbp]\centering
    \includegraphics[width=0.47\textwidth]{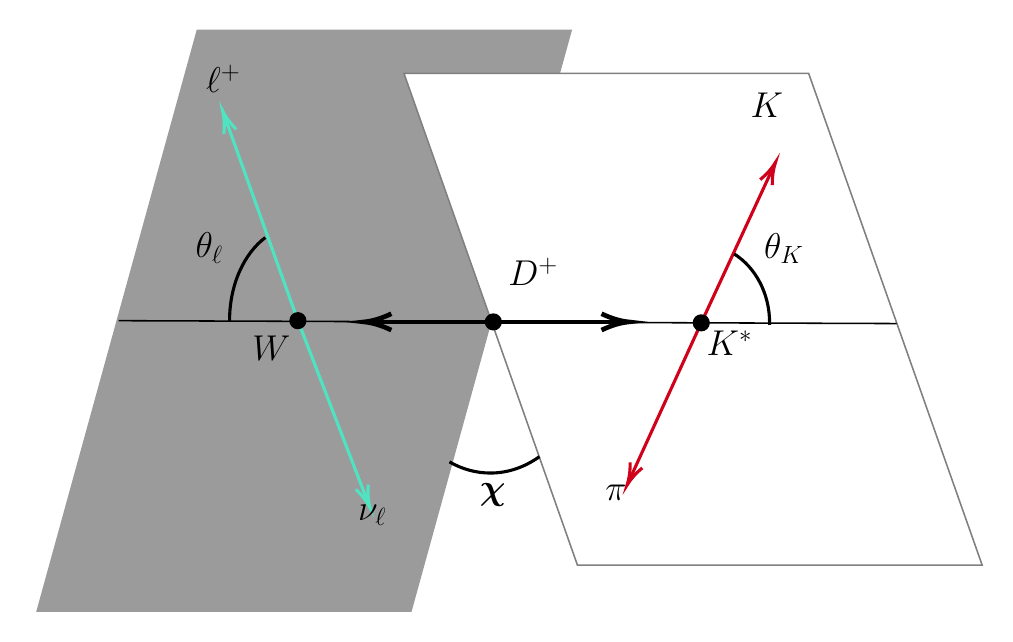}
    \caption{Definition of the angular variables in the $D^+\to \bar{K} \pi \ell^+\nu_\ell$ decay.}
    \label{fig:Kine}
\end{figure}

The variable $\cos \theta_\ell$ ($\cos \theta_K$) is the cosine of the angle between the momentum of the charged lepton (neutral kaon) in the rest frame of the $\ell\nu$ ($K\pi$) system with respect to the $\ell\nu$ ($K\pi$) flight direction as seen from the rest frame of the candidate~\cite{ke4pipiphase} (see Fig.~\ref{fig:Kine}):
\begin{equation}
\begin{aligned}
\cos{\theta_\ell} &= \vec{e}_{\ell\nu} \cdot \vec{e}_\ell, \\
\cos{\theta_K} &= \vec{e}_{K\pi} \cdot \vec{e}_K. 
\end{aligned}
\end{equation}
Here, $\vec{e}_{\ell\nu}$ ($\vec{e}_{K\pi}$) is the unit vector along the momentum of the $\ell\nu$ ($K\pi$) system in the rest frame of the $W^*$ ($K^*$) and $\vec{e}_\ell$ ($\vec{e}_K$) is the unit vector along the momentum of the charged lepton (neutral kaon) in the $\ell\nu$ ($K\pi$) rest frame. The angle $\chi$ is the angle between the two decay planes of the $\ell\nu$ and $K\pi$ systems, defined by:    
\begin{equation}
\begin{aligned}   
\cos{\chi} &= \vec{n}_{\ell\nu} \cdot \vec{n}_{K\pi}, \\
\sin{\chi} &= [\vec{n}_{\ell\nu} \times \vec{n}_{K\pi}] \cdot \vec{e}_{K\pi}, 
\end{aligned}
\end{equation}
where $\vec{n}_{\ell\nu}=\vec{e}_\ell\times\vec{e}_\nu$ ($\vec{n}_{K\pi}=\vec{e}_K\times\vec{e}_\pi$) is defined as the unit vector perpendicular to the decay plane spanned by the two leptons (or the two hadrons). The angles are defined in $-1 \leq \cos \theta_K \leq 1$, $ -1 \leq \cos \theta_\ell \leq 1$ and $-\pi \leq \chi \leq \pi$. The same definition of the angles with Ref. \cite{BESanfenfen}. When analyzing $D^-$ decays, the sign of $\chi$ is reversed.
 
\section{Formalism of the Angular Coefficients Asymmetries}
The eight coefficients $\mathcal{I}_{2-9}$ can be obtained by defining piece-wise angular-integration ranges in $\chi$ and $\cos\theta_\ell$ as
\begin{equation}\label{eq:integralcosLchi}
    \begin{aligned}
    & \mathcal{I}_2= \int_{-\pi}^\pi d\chi\left[\int_{-1}^{-0.5} d\cos\theta_\ell+\int_{0.5}^1 d\cos\theta_\ell-\int_{-0.5}^{0.5} d\cos\theta_\ell\right] \,\frac{d^5 \Gamma}{dq^2 \, dm^2 \, d\Omega}, \\
    & \mathcal{I}_3= \left[\int_{-\pi}^{-\frac{3 \pi}{4}} d\chi+\int_{-\frac{\pi}{4}}^{\frac{\pi}{4}} d\chi+\int_{\frac{3 \pi}{4}}^\pi d\chi-\int_{-\frac{3 \pi}{4}}^{-\frac{\pi}{4}} d\chi-\int_{\frac{\pi}{4}}^{\frac{3 \pi}{4}} d\chi\right] \int_{-1}^1 d\cos\theta_\ell \,\frac{d^5 \Gamma}{dq^2 \,dm^2 \, d\Omega}, \\
    & \mathcal{I}_4= \left[\int_{-\frac{\pi}{2}}^{\frac{\pi}{2}} d\chi-\int_{-\pi}^{-\frac{\pi}{2}} d\chi-\int_{\frac{\pi}{2}}^\pi d\chi\right]\left[\int_0^1 d\cos\theta_\ell-\int_{-1}^0 d\cos\theta_\ell\right] \,\frac{d^5 \Gamma}{dq^2 \,dm^2 \, d\Omega}, \\
    & \mathcal{I}_5= \left[\int_{-\frac{\pi}{2}}^{\frac{\pi}{2}} d\chi-\int_{-\pi}^{-\frac{\pi}{2}} d\chi-\int_{\frac{\pi}{2}}^\pi d\chi\right] \int_{-1}^1 d\cos\theta_\ell \,\frac{d^5 \Gamma}{dq^2 \,dm^2 \, d\Omega}, \\
    & \mathcal{I}_6= \int_{-\pi}^\pi d\chi\left[\int_0^1 d\cos\theta_\ell-\int_{-1}^0 d\cos\theta_\ell\right] \,\frac{d^5 \Gamma}{dq^2 \,dm^2 \, d\Omega} \\
    & \mathcal{I}_7= \left[\int_0^\pi d\chi-\int_{-\pi}^0 d\chi\right] \int_{-1}^1 d\cos\theta_\ell \,\frac{d^5 \Gamma}{dq^2 \,dm^2 \, d\Omega}, \\
    & \mathcal{I}_8= \left[\int_0^\pi d\chi-\int_{-\pi}^0 d\chi\right]\left[\int_0^1 d\cos\theta_\ell-\int_{-1}^0 d\cos\theta_\ell\right] \,\frac{d^5 \Gamma}{dq^2 \,dm^2 \, d\Omega}, \\
    & \mathcal{I}_9= \left[\int_{-\pi}^{-\frac{\pi}{2}} d\chi+\int_0^{\frac{\pi}{2}} d\chi-\int_{-\frac{\pi}{2}}^0 d\chi-\int_{\frac{\pi}{2}}^\pi d\chi\right] \int_{-1}^1 d\cos\theta_\ell \,\frac{d^5 \Gamma}{dq^2 \,dm^2 \, d\Omega}.
    \end{aligned}
\end{equation}
 
Experimentally, the observables $\left\langle \mathcal{I}_i \right\rangle$ are determined by measuring the decay-rate asymmetries of the data split by angular tags, $T_i$, defined according to the piece-wise integration of Eqs.~\ref{eq:integralcosLchi} above and Eq.~4 in the main text as 
\begin{equation}
    \left\langle \mathcal{I}_i \right\rangle=\frac{1}{\Gamma} \left[\Gamma\left(T_i=\text{true}\right)-\Gamma\left(T_i=\text{false}\right)\right],
\end{equation}
with 
\begin{equation}
    \begin{aligned}
    & T_2=\left|\cos \theta_\ell\right|>0.5, \\
    & T_3=\cos 2 \chi>0, \\
    & T_4=\left(\sin 2 \theta_\ell>0 ; \cos \chi>0 ; \cos \theta_K>0\right) \text { or }\left(\sin 2 \theta_\ell>0 ; \cos \chi<0 ; \cos \theta_K<0\right) \\
    & \quad \text { or }\left(\sin 2 \theta_\ell<0 ; \cos \chi<0 ; \cos \theta_K>0\right) \text { or }\left(\sin 2 \theta_\ell<0 ; \cos \chi>0 ; \cos \theta_K<0\right), \\
    & T_5=\left(\cos \chi>0 ; \cos \theta_K>0\right) \text { or }\left(\cos \chi<0 ; \cos \theta_K<0\right), \\
    & T_6=\cos \theta_\ell>0, \\
    & T_7=\left(\sin \chi>0 ; \cos \theta_K>0\right) \text { or }\left(\sin \chi<0 ; \cos \theta_K<0\right), \\
    & T_8=\left(\sin 2 \theta_\ell>0 ; \sin \chi>0 ; \cos \theta_K>0\right) \text { or }\left(\sin 2 \theta_\ell>0 ; \sin \chi<0 ; \cos \theta_K<0\right) \\
    & \quad \text { or }\left(\sin 2 \theta_\ell<0 ; \sin \chi<0 ; \cos \theta_K>0\right) \text { or }\left(\sin 2 \theta_\ell<0 ; \sin \chi>0 ; \cos \theta_K<0\right), \\
    & T_9=\sin 2 \chi>0.
    \end{aligned}
\end{equation}

The full set of measured $CP$ asymmetries and averaged observables are listed in Table~\ref{tab:asyAS}. The $CP$ asymmetries in the decay rates $A_{CP}$, are listed in Table~\ref{tab:asyAcp}.

\begin{table}[htbp]\centering
    \caption{Angular observables $\left\langle A_i \right\rangle$ and $\left\langle S_i \right\rangle$ (in \%) for $D^+ \to \bar{K}^{*0}\ell^+\nu_\ell$ decay in different $q^2$ regions with $0.8<m_{\bar{K}^{*0}}<1.0$ GeV$^2/c^4$, where the first uncertainties are statistical and the second are systematic.}
    \resizebox{1.0\textwidth}{!}{
    \begin{tabular}{ccccccccc}
        \hline\hline
        $q^2$ (GeV$^2/c^4$) & $\langle A_2\rangle$ & $\langle A_3\rangle$ & $\langle A_4\rangle$ & $\langle A_5\rangle$ & $\langle A_6\rangle$ & $\langle A_7\rangle$ & $\langle A_8\rangle$ & $\langle A_9\rangle$ \\
        \hline
        \hline \multicolumn{9}{c}{$D^+ \to \bar{K}^{*0}e^+\nu_e$} \\\hline
        $[q^2_\text{min},0.20]$  &  $ 6.7\pm4.6\pm0.1$ & $-2.8\pm3.2\pm0.4$ & $ 1.9\pm4.0\pm0.4$ & $ 2.2\pm3.2\pm0.3$ & $ 2.2\pm4.0\pm0.4$ & $ 1.9\pm3.1\pm0.2$ & $-5.6\pm4.0\pm0.2$ & $-1.4\pm3.2\pm0.2$ \\
        $[0.20,0.40]$            &  $ 1.6\pm3.1\pm0.1$ & $ 0.1\pm2.9\pm0.1$ & $-4.1\pm3.0\pm0.1$ & $-1.4\pm2.9\pm0.2$ & $ 7.2\pm3.0\pm0.1$ & $-1.0\pm2.9\pm0.2$ & $-0.7\pm3.0\pm0.3$ & $ 0.4\pm2.9\pm0.1$ \\
        $[0.40,0.60]$            &  $-2.9\pm2.6\pm0.1$ & $-6.0\pm2.7\pm0.2$ & $-5.4\pm3.1\pm0.1$ & $ 2.8\pm3.0\pm0.2$ & $-2.1\pm2.9\pm0.2$ & $-1.3\pm3.0\pm0.1$ & $ 0.3\pm3.1\pm0.1$ & $ 0.2\pm3.1\pm0.1$ \\
        $[0.60,0.80]$            &  $-0.5\pm3.3\pm0.3$ & $ 3.4\pm3.4\pm0.4$ & $ 2.8\pm3.3\pm0.3$ & $ 0.0\pm3.3\pm0.2$ & $ 0.6\pm3.2\pm0.3$ & $ 0.1\pm3.2\pm0.0$ & $ 1.0\pm3.3\pm0.6$ & $-2.9\pm3.4\pm0.1$ \\
        $[0.80,q^2_\text{max}]$  &  $ 1.8\pm4.7\pm0.5$ & $ 3.1\pm4.9\pm0.3$ & $-2.5\pm5.1\pm0.4$ & $ 1.0\pm5.0\pm1.1$ & $ 4.6\pm4.8\pm0.1$ & $ 4.0\pm4.9\pm0.1$ & $-2.5\pm5.1\pm0.6$ & $ 6.6\pm4.9\pm0.3$ \\
        Full range               &  $ 0.4\pm1.4\pm0.1$ & $-1.1\pm1.3\pm0.1$ & $-1.6\pm1.4\pm0.1$ & $ 0.6\pm1.3\pm0.1$ & $ 1.9\pm1.3\pm0.1$ & $ 0.2\pm1.2\pm0.1$ & $-1.0\pm1.4\pm0.1$ & $ 0.3\pm1.3\pm0.0$ \\
        \hline
        \hline \multicolumn{9}{c}{$D^+ \to \bar{K}^{*0}\mu^+\nu_\mu$} \\\hline
        $[q^2_\text{min},0.20]$  & $ -6.2\pm5.1\pm1.3$ & $ 2.8\pm4.5\pm0.3$ & $-4.9\pm5.3\pm2.1$ & $ -4.2\pm4.4\pm0.3$ & $-3.5\pm7.8\pm0.6$ & $  6.0\pm4.5\pm0.7$ & $-9.9\pm5.2\pm0.3$ & $ 7.0\pm4.7\pm0.4$ \\
        $[0.20,0.40]$            & $ -5.8\pm4.0\pm0.4$ & $ 0.2\pm4.0\pm0.2$ & $-8.1\pm4.0\pm0.2$ & $  0.5\pm3.9\pm0.5$ & $-5.6\pm4.5\pm0.2$ & $  4.8\pm3.8\pm0.7$ & $-0.1\pm3.9\pm0.2$ & $-3.8\pm3.9\pm0.9$ \\
        $[0.40,0.60]$            & $  5.7\pm4.0\pm0.5$ & $-0.3\pm4.1\pm0.3$ & $-0.5\pm4.1\pm0.4$ & $  1.1\pm4.0\pm0.1$ & $ 4.0\pm4.2\pm0.1$ & $-11.3\pm4.0\pm0.2$ & $-2.5\pm4.2\pm0.2$ & $ 1.9\pm4.0\pm0.2$ \\
        $[0.60,0.80]$            & $  3.7\pm5.1\pm0.3$ & $ 7.4\pm5.3\pm0.5$ & $-0.9\pm5.1\pm1.2$ & $-10.0\pm5.6\pm0.2$ & $ 2.7\pm5.6\pm0.8$ & $  3.4\pm5.1\pm0.3$ & $ 4.7\pm5.2\pm0.7$ & $10.7\pm5.0\pm0.5$ \\
        $[0.80,q^2_\text{max}]$  & $-10.3\pm7.9\pm0.4$ & $-8.5\pm7.3\pm3.9$ & $-3.5\pm7.7\pm1.2$ & $  3.7\pm7.3\pm4.3$ & $-6.2\pm6.5\pm0.0$ & $-11.7\pm7.5\pm0.4$ & $ 2.5\pm7.3\pm0.0$ & $-4.1\pm8.4\pm1.7$ \\
        Full range               & $ -1.7\pm2.1\pm0.5$ & $ 1.7\pm1.9\pm0.2$ & $-3.2\pm2.1\pm0.1$ & $ -2.1\pm2.0\pm0.2$ & $ 0.4\pm2.2\pm0.5$ & $ -0.7\pm2.0\pm0.4$ & $-0.9\pm2.1\pm0.2$ & $ 2.5\pm2.0\pm0.4$ \\
        \hline\hline
        $q^2$ (GeV$^2/c^4$) & $\langle S_2\rangle$ & $\langle S_3\rangle$ & $\langle S_4\rangle$ & $\langle S_5\rangle$ & $\langle S_6\rangle$ & $\langle S_7\rangle$ & $\langle S_8\rangle$ & $\langle S_9\rangle$ \\
        \hline
        \hline \multicolumn{9}{c}{$D^+ \to \bar{K}^{*0}e^+\nu_e$} \\\hline
        $[q^2_\text{min},0.20]$  & $-31.7\pm4.9\pm1.1$ & $-10.1\pm3.3\pm0.2$ & $15.4\pm4.0\pm0.6$ & $-18.1\pm3.3\pm1.3$ & $-14.7\pm4.0\pm2.9$ & $ 0.2\pm3.1\pm0.1$ & $-0.2\pm3.9\pm0.2$ & $ 5.2\pm3.2\pm0.4$ \\
        $[0.20,0.40]$            & $-14.0\pm3.2\pm0.8$ & $ -7.0\pm2.9\pm0.2$ & $19.6\pm3.1\pm0.6$ & $-19.7\pm2.9\pm0.2$ & $-17.3\pm3.0\pm0.7$ & $ 7.1\pm2.9\pm0.2$ & $-1.2\pm3.0\pm0.5$ & $ 0.7\pm2.9\pm0.2$ \\
        $[0.40,0.60]$            & $ -7.5\pm2.6\pm0.7$ & $-15.9\pm2.8\pm0.2$ & $14.2\pm3.1\pm0.3$ & $-15.6\pm3.0\pm0.4$ & $-25.7\pm3.0\pm0.4$ & $-1.7\pm3.0\pm0.1$ & $-1.3\pm3.1\pm0.0$ & $ 0.8\pm3.1\pm0.3$ \\
        $[0.60,0.80]$            & $ -8.9\pm3.3\pm0.4$ & $-14.3\pm3.5\pm0.2$ & $22.5\pm3.4\pm0.5$ & $-15.1\pm3.4\pm0.2$ & $-19.6\pm3.3\pm0.2$ & $ 2.5\pm3.2\pm0.1$ & $-0.6\pm3.3\pm0.3$ & $ 1.3\pm3.4\pm0.0$ \\
        $[0.80,q^2_\text{max}]$  & $  0.8\pm4.7\pm1.0$ & $-14.2\pm5.0\pm0.4$ & $26.2\pm5.2\pm0.4$ & $ -7.2\pm5.0\pm1.1$ & $-17.9\pm5.0\pm0.1$ & $ 0.7\pm5.0\pm0.3$ & $-3.1\pm5.1\pm0.6$ & $-7.5\pm4.9\pm0.1$ \\
        Full range               & $-13.2\pm1.4\pm0.2$ & $-11.9\pm1.4\pm0.1$ & $18.9\pm1.4\pm0.1$ & $-16.1\pm1.3\pm0.1$ & $-18.7\pm1.4\pm0.6$ & $ 2.3\pm1.2\pm0.1$ & $-1.1\pm1.4\pm0.0$ & $ 0.4\pm1.3\pm0.0$ \\
        \hline
        \hline \multicolumn{9}{c}{$D^+ \to \bar{K}^{*0}\mu^+\nu_\mu$} \\\hline
        $[q^2_\text{min},0.20]$  & $-26.1\pm5.5\pm1.1$ & $  1.0\pm4.5\pm1.0$ & $ 7.4\pm5.3\pm1.6$ & $-10.6\pm4.4\pm0.2$ & $ -6.6\pm7.4\pm1.2$ & $-9.0\pm4.5\pm0.5$ & $10.3\pm5.2\pm0.7$ & $ 2.7\pm4.7\pm0.6$ \\
        $[0.20,0.40]$            & $ -9.0\pm4.0\pm0.2$ & $ -4.4\pm4.0\pm0.3$ & $23.8\pm4.1\pm0.5$ & $-17.8\pm4.0\pm0.7$ & $-26.3\pm4.3\pm0.8$ & $ 3.2\pm3.8\pm0.1$ & $ 0.9\pm3.9\pm0.2$ & $ 3.4\pm3.9\pm0.2$ \\
        $[0.40,0.60]$            & $ -9.8\pm4.1\pm0.1$ & $-13.9\pm4.1\pm0.2$ & $22.2\pm4.2\pm0.9$ & $-18.5\pm4.0\pm0.3$ & $-28.7\pm4.0\pm0.6$ & $ 0.3\pm4.0\pm0.3$ & $ 1.6\pm4.2\pm0.2$ & $-2.9\pm4.0\pm0.1$ \\
        $[0.60,0.80]$            & $ -4.3\pm5.1\pm0.3$ & $-10.0\pm5.2\pm0.2$ & $20.3\pm5.2\pm2.2$ & $-15.6\pm5.3\pm0.3$ & $-19.3\pm5.6\pm0.3$ & $-4.6\pm5.1\pm0.0$ & $-8.1\pm5.3\pm0.1$ & $ 7.5\pm5.0\pm0.4$ \\
        $[0.80,q^2_\text{max}]$  & $ 10.4\pm8.3\pm0.4$ & $-13.1\pm7.3\pm4.3$ & $15.9\pm7.9\pm0.3$ & $ -8.9\pm7.3\pm4.5$ & $-15.0\pm6.4\pm0.8$ & $ 1.7\pm7.5\pm0.4$ & $-2.5\pm7.3\pm0.3$ & $ 6.7\pm8.7\pm1.9$ \\
        Full range               & $ -8.9\pm2.1\pm0.3$ & $ -8.5\pm2.0\pm0.2$ & $19.6\pm2.1\pm0.1$ & $-15.5\pm2.0\pm0.4$ & $-23.4\pm2.3\pm0.3$ & $-1.9\pm2.0\pm0.3$ & $ 0.6\pm2.1\pm0.1$ & $ 1.9\pm2.0\pm0.3$ \\
        \hline\hline
    \end{tabular}
    }
    \label{tab:asyAS}
\end{table}

\begin{table}[htbp]\centering
    \caption{Observables $A_{CP}$ (in \%) for $D^+ \to \bar{K}^{*0}\ell^+\nu_\ell$ decay in different $q^2$ regions with $0.8<m_{\bar{K}^{*0}}<1.0$ GeV$^2/c^4$, where the first uncertainties are statistical and the second are systematic.}
    \begin{tabular}{ccc}
        \hline\hline
        $q^2$ (GeV$^2/c^4$) & $D^+ \to \bar{K}^{*0}e^+\nu_e$ & $D^+ \to \bar{K}^{*0}\mu^+\nu_\mu$  \\
        \hline
        $[q^2_\text{min},0.20]$  & $ 2.4\pm2.9\pm0.1$ & $ 3.0\pm4.4\pm0.2$ \\
        $[0.20,0.40]$            & $ 0.5\pm2.8\pm0.1$ & $-2.8\pm3.8\pm0.2$ \\
        $[0.40,0.60]$            & $ 0.4\pm3.0\pm0.1$ & $-6.9\pm3.9\pm0.1$ \\
        $[0.60,0.80]$            & $-0.4\pm3.2\pm0.0$ & $-1.7\pm5.0\pm0.1$ \\
        $[0.80,q^2_\text{max}]$  & $ 3.2\pm4.8\pm0.1$ & $-5.5\pm7.0\pm1.5$ \\
        Full range               & $ 1.6\pm1.2\pm0.4$ & $-2.3\pm1.9\pm0.2$ \\
        \hline\hline
    \end{tabular}
    \label{tab:asyAcp}
\end{table}

\clearpage

\section{Helicity amplitude of $D^+\to \bar{K}^{*0}\ell\nu_\ell$}

The angular coefficients can be expanded in terms of four form factors $\mathcal{F}_{1,2,3,4}$ as
\begin{equation}\label{eq:definition_decay_intensity}
    \begin{aligned}
    \mathcal{I}_1= & \frac{\beta_\ell}{4} \left[\left(1+\frac{m^2_\ell}{q^2}\right)|\mathcal{F}_1|^2+\frac{3}{2}\left(1+\frac{m^2_\ell}{3q^2}\right) \sin ^2 \theta_K\left(|\mathcal{F}_2|^2+|\mathcal{F}_3|^2\right) +\frac{2m^2_\ell}{q^2}|\mathcal{F}_4|^2\right], \\
    \mathcal{I}_2= & -\frac{\beta_\ell^2}{4}\left[|\mathcal{F}_1|^2-\frac{1}{2} \sin ^2 \theta_K \left(|\mathcal{F}_2|^2+|\mathcal{F}_3|^2\right)\right], \\
    \mathcal{I}_3= & -\frac{\beta_\ell^2}{4}\left[\left(|\mathcal{F}_2|^2-|\mathcal{F}_3|^2\right)\right]\sin ^2 \theta_K, \\
    \mathcal{I}_4= & \frac{\beta_\ell^2}{2} \operatorname{Re}\left(\mathcal{F}_1 \mathcal{F}_2^{*}\right)\sin \theta_K, \\
    \mathcal{I}_5= & \beta_\ell\operatorname{Re}\left[\mathcal{F}_1 \mathcal{F}_3^{*}+ \frac{m^2_\ell}{q^2} \mathcal{F}_4\mathcal{F}_2^{*}\right]\sin \theta_K, \\
    \mathcal{I}_6= & \beta_\ell \operatorname{Re}\left[\mathcal{F}_2 \mathcal{F}_3^{*}\sin^2 \theta_K -\frac{m^2_\ell}{q^2}\mathcal{F}_1 \mathcal{F}_4^{*}\right], \\
    \mathcal{I}_7= & \beta_\ell \operatorname{Im}\left[\left(\mathcal{F}_1 \mathcal{F}_2^{*}\right)+\frac{m^2_\ell}{q^2}\mathcal{F}_4\mathcal{F}_3^{*}\right]\sin \theta_K, \\
    \mathcal{I}_8= & \frac{\beta_\ell^2}{2} \operatorname{Im}\left(\mathcal{F}_1 \mathcal{F}_3^{*}\right)\sin \theta_K,\\
    \mathcal{I}_9= & -\frac{\beta_\ell^2}{2} \operatorname{Im}\left(\mathcal{F}_2 \mathcal{F}_3^{*}\right)\sin ^2 \theta_K.\\
    \end{aligned}
\end{equation}

The form factors $\mathcal{F}_{1,2,3,4}$ can be expanded into partial waves to show their explicit dependence on $\theta_{K}$. If only
$S$- and $P$-waves are kept, one obtains:
\begin{equation}
    \begin{aligned}
        & \mathcal{F}_1=\sum_{l=0}^{\infty} \mathcal{F}_{1, l} \, P_l\left(\cos \theta_K\right) = \mathcal{F}_{10} + \mathcal{F}_{11} \cos \theta_K,\\
        & \mathcal{F}_2=\sum_{l=1}^{\infty} \frac{1}{\sqrt{l(l+1)}} \, \mathcal{F}_{2, l} \, \frac{d}{d \cos \theta_K} \, P_l\left(\cos \theta_K\right)=\frac{1}{\sqrt{2}} \, \mathcal{F}_{21},\\
        & \mathcal{F}_3=\sum_{l=1}^{\infty} \frac{1}{\sqrt{l(l+1)}} \, \mathcal{F}_{3, l} \, \frac{d}{d \cos \theta_K} \, P_l\left(\cos \theta_K\right)=\frac{1}{\sqrt{2}} \, \mathcal{F}_{31},\\
        & \mathcal{F}_4=\sum_{l=0}^{\infty} \mathcal{F}_{4, l} \, P_l\left(\cos \theta_K\right) = \mathcal{F}_{40} + \mathcal{F}_{41} \cos \theta_K.
    \end{aligned}
\end{equation}

By comparing expressions given in Refs.~\cite{bl4-dl4,Richman:1995wm} it is possible to relate $\mathcal{F}_{i1},~i=1,2,3,4$ with the helicity form factors $H_{0,\pm,t}$:
\begin{equation}
    \begin{aligned}
    \mathcal{F}_{11} = & 2\sqrt{2} \,\alpha q\, H_{0} \; \mathcal{A}(m), \\
    \mathcal{F}_{21} = & 2\alpha q\,(H_{+} + H_{-}) \; \mathcal{A}(m), \\
    \mathcal{F}_{31} = & 2\alpha q\,(H_{+} - H_{-}) \; \mathcal{A}(m), \\
    \mathcal{F}_{41} = & 2\sqrt{2} \,\alpha q\, H_{t} \; \mathcal{A}(m),
    \label{eq:form_factor_P}
    \end{aligned}
\end{equation}
where $\alpha = \sqrt{3\pi \mathcal{B}_{\bar K^{*0}}/(p^*_0\Gamma_0)}$ and the $\mathcal{F}_i$ depend on the mass-dependent amplitude $\mathcal{A}(m)$.  
Here, $\mathcal{B}_{\bar{K}^{*0}}=\mathcal{B}(\bar{K}^{*0}\to \bar{K}^0\pi^0)=1/3$. The helicity form factors can then be related to two axial-vector form factors $A_{0,1}(q^2)$ and one vector form factor $V(q^2)$:
\begin{equation}
    \begin{aligned}
    H_{0}(q^2, m^2) & = \frac{1}{2mq}\left[(m_D^2-m^2-q^2)(m_D+m)A_{1}(q^2)-4\frac{m_D^2p_{K_S^0\pi^0}^2}{m_D+m}A_{2}(q^2)\right], \\
    H_{\pm}(q^2, m^2) & = (m_D+m)A_{1}(q^2)\mp\frac{2m_Dp_{K_S^0\pi^0}}{(m_D+m)}V(q^2),\\
    H_{t}(q^2, m^2) & = \frac{2m_D p_{K_S^0\pi^0}}{\sqrt{q^2}} \left[\frac{m_D+m}{2 m} A_1(q^2)-\frac{m_D-m}{2 m} A_2(q^2)\right].\\
    \label{eq:helicity}
    \end{aligned}
\end{equation}

For the $q^2$ dependence we use a single-pole dominance parametrization (SPD):
\begin{equation}\label{eq:spd}
    \begin{aligned}
    V(q^2)=\frac{V(0)}{1-q^2/m_V^2},\quad A_{0,1,2}(q^2)=\frac{A_{0,1,2}(0)}{1-q^2/m_A^2},\\
    \end{aligned}
\end{equation}
where the pole masses $m_V$ and $m_A$ are fixed at 1.81GeV/$c^2$ and 2.61 GeV/$c^2$~\cite{BESanfenfen}, respectively. For the mass dependence for $\bar{K}^{*0}$, we adopt a Breit-Wigner distribution with a mass-dependent width as
\begin{equation}\label{eq:Am}
    \mathcal{A}(m) = \frac{m_{0} \Gamma_{0} F_1(m)}{m_{0}^{2} - m^{2} - im_{0} \Gamma(m)},
\end{equation}
where $m_0$ and $\Gamma_0$ denote its pole mass and total width, respectively. The momentum-dependent width $\Gamma(m)$ is expressed as:
\begin{equation}\label{eq:FL}
    \Gamma(m) = \Gamma_{0}\frac{p^*}{p^*_0}\frac{m_0}{m}F_1^2, \quad F_1 =\frac{p^*}{p^*_0}\frac{B(p^*)}{B(p^*_0)},
\end{equation}
here $p^*$ is the momentum of the $K_S^0$ in the $K_S^0\pi^0$ rest frame, and $p^*_0$ is the corresponding value taken at the pole mass of the resonance. $B$ is the Blatt-Weisskopf damping factor~\cite{rBW}, defined as $B = 1/\sqrt{1 + r_{BW}^{2}p^{*2}}$, the barrier factor, $r_{BW}$, related to the meson effective radii is fixed to 3.07 GeV$^{-1}$~\cite{BESanfenfen}.

The $S$-wave related form factors $\mathcal{F}_{10}$ and $\mathcal{F}_{40}$ are expressed as
\begin{equation}
    \begin{aligned}
        \mathcal{F}_{10}=\,&p_{K_S^0\pi^0}m_{D} \, \frac{1}{1-q^2/m_A^2} \, \mathcal{A}_{S}(m), \\
        \mathcal{F}_{40}=\,&\frac{q^2}{1-q^2/m_A^2} \, \mathcal{A}_{S}(m).
    \end{aligned}
    \label{eq:form_factor_S}
\end{equation}

We consider the $S$-wave amplitude $\mathcal{A}_{S}(m)$ as the combination of a non-resonant background and $\bar{K}^{*}_{0}(1430)$. According to Watson's theorem~\cite{Swavetheory}, for the same isospin and angular momentum, the phase measured in the $K_S^0\pi^0$ elastic scattering and in a decay channel are equal in the elastic regime. So we take the phase parameterization used in the LASS scattering experiment for our $S$-wave amplitude~\cite{Swavelass}:
\begin{equation}
    \begin{aligned}
        \cot(\delta_\text{BG}^{1/2})=\frac{1}{a_{S, \text{BG}}^{1/2}p^*}+\frac{b_{S, \text{BG}}^{1/2}p^*}{2}, \quad 
        \cot(\delta_{K^{*}_{0}(1430)})=\frac{m^2_{K^{*}_{0}(1430)}-m^2}{m_{K^{*}_{0}(1430)}\Gamma_{K^{*}_{0}(1430)}(m)}, \quad 
        \delta(m)=\delta_\text{BG}^{1/2} + \delta_{K^{*}_{0}(1430)},
    \end{aligned}
\end{equation}
where $\delta_\text{BG}^{1/2} $ is the scattering length and $b_{S, \text{BG}}^{1/2}$ is the effective range; $\delta_\text{BG}^{1/2} $ is determined by the fit, and $b_{S, \text{BG}}^{1/2}$ is fixed at $-0.81$~\cite{BESanfenfen}.

For the magnitude of the $\mathcal{A}_{S}(m)$, we take the formalism from the BaBar experiment~\cite{BaBar2010}. This entails a piecewise function for $m$ below and above the $\mathcal{A}_{S}(m)$ pole mass value:
\begin{equation}\centering
    \begin{aligned}
        \mathcal{A}_{S}(m)=r_{S}P(m)e^{i\delta(m)}, \quad m < m_{K^{*}_{0}(1430)};\\
        \mathcal{A}_{S}(m)=r_{S}P(m_{K^{*}_{0}(1430)})\sqrt{\frac{(m_{K^{*}_{0}(1430)}\Gamma_{K^{*}_{0}(1430)})^2}{(m^2_{K^{*}_{0}(1430)}-m^2)^2+(m_{K^{*}_{0}(1430)}\Gamma_{K^{*}_{0}(1430)})^2}}e^{i\delta(m)}, \quad m > m_{K^{*}_{0}(1430)}.
    \end{aligned}
\end{equation}
In these expressions, $P(m)=1+r^{(1)}_S x$, $x = \sqrt{(\frac{m}{m_{K_S^0}+m_{\pi^0}})-1}$. The magnitude $|\mathcal{A}_s(m)|$ is assumed to have a linear variation versus $x$ below the pole mass while having a Breit-Wigner form above it. The dimension less coefficient $r^{(1)}_S$ is fixed at $0.08$~\cite{BESanfenfen}, and the relative intensity $r_S$ will be determined by the fit.

For one candidate event, the probability density function (PDF) is expressed as:
\begin{equation}\label{eq:pdf}
    \text{PDF}(\xi, \eta) = (1-f_b) \mathcal{S} + f_b \mathcal{B} = (1-f_b) \frac{\omega(\xi, \eta) \epsilon(\xi) R_4(\xi)}{\int d\xi \omega(\xi, \eta) \epsilon(\xi) R_4(\xi)} + f_b \frac{B(\xi) R_4(\xi)}{\int d\xi B(\xi) R_4(\xi)},
\end{equation}
where $\xi$ represents the kinematic variables of the event and $\eta$ represents the model parameters. Here, $\omega(\xi,\eta)$ is the decay intensity defined inside the parentheses in Eq.~1 in main text, ${B}(\xi)$ is the background function and $\epsilon(\xi)$ is the reconstruction efficiency for final state $\xi$.

The likelihood is calculated as the product of the probabilities for all $N$ observed events:
\begin{equation}
    \begin{aligned}
        \mathcal{L} = \prod_{i=1}^{N} \text{PDF}(\xi_i, \eta) = \prod_{i=1}^{N} \epsilon(\xi_i)R_4(\xi_i) \left[(1-f_b) \frac{\omega(\xi_i,\eta)}{\int d\xi_i \omega(\xi_i,\eta) \epsilon(\xi_i)R_4(\xi_i)} + f_b \frac{B_\epsilon(\xi_i)}{\int d\xi_i B_\epsilon(\xi_i) \epsilon(\xi_i)R_4(\xi_i)}\right],\\
        -\ln \mathcal{L} =
        -\sum_{i=1}^{N}\ln(\epsilon(\xi_i)R_4(\xi_i))
        -\sum_{i=1}^{N}\ln \left[(1-f_b) \frac{\omega(\xi_i,\eta)}{\int d\xi_i \omega(\xi_i,\eta) \epsilon(\xi_i)R_4(\xi_i)} + f_b \frac{B_\epsilon(\xi_i)}{\int d \xi_i B_\epsilon(\xi_i)\epsilon(\xi_i)R_4(\xi_i)}\right].\\
    \end{aligned}
\end{equation}
where $B_\epsilon(\xi)$ represents the background model, corrected for acceptance and phase space. The constant term, dependent on event selection efficiency, remains unchanged during the fit. So actually we only compute the second term for optimization. The normalization integral terms is calculate by making MC integration with signal MC sample~\cite{pwa-MCintegration}. The background PDF is modeled with the inclusive MC sample using the XGBoost algorithm~\cite{xgboost}.

\section{Single Tag information}

\begin{figure}[htbp]\centering
    \includegraphics[width=0.6\textwidth]{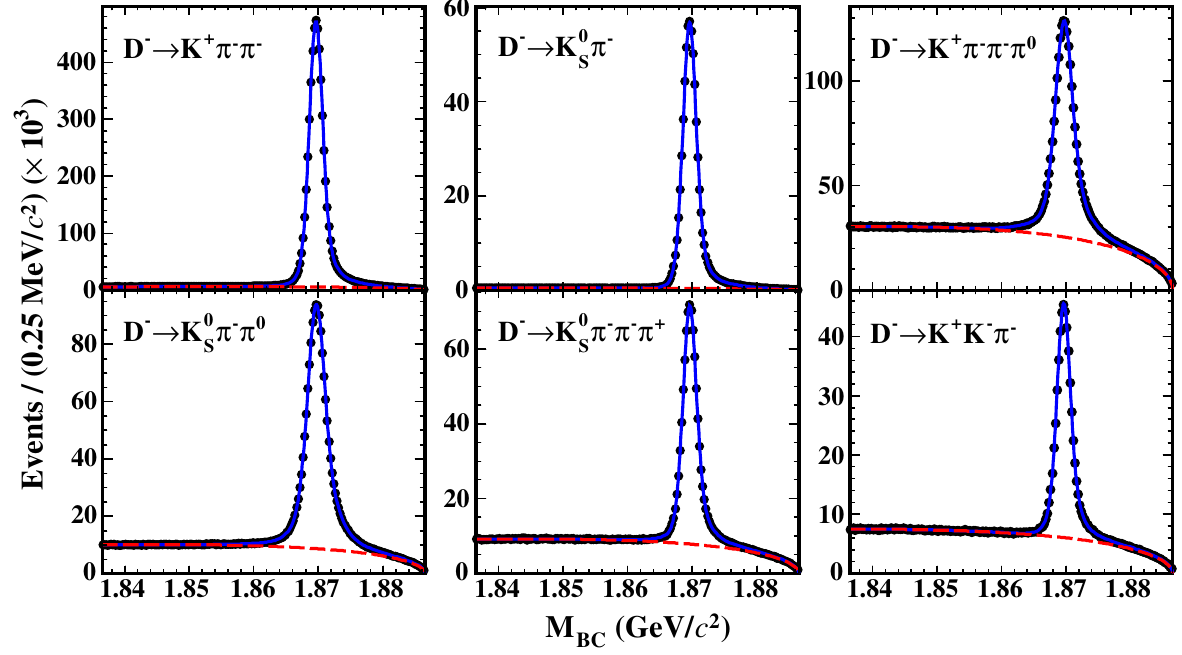}
    \caption{Fits to the $M_{\rm BC}$ distributions of tag channels. The dots with error bars are data, the blue solid curves are the total fits, the red dashed lines indicate the background contributions. The pair of red arrows indicate the $M_{\rm BC}$ signal window.}
    \label{fig:STfit}
\end{figure}

Figure~\ref{fig:STfit} shows the fits to the $M_{\rm BC}$ distributions of the selected ST candidates in data for different tag modes. Table~\ref{tab:STyield} summaries the requirements on $\Delta E$, the ST yields in data, and the ST efficiencies.

\begin{table}[htbp]\centering
    \caption{The $\Delta E$ requirements, ST $D^-$ yields in data, and ST efficiency.}
    \begin{tabular}{lccc}
        \hline\hline
        Tag mode                          & $\Delta E$(GeV)  & $N^{i}_{\rm data}$~($\times 10^{4}$) & $\bar{\epsilon}^{i}_{\rm tag} (\%)$ \\\hline
        $D^- \to K^+ \pi^- \pi^-$         & $(-0.025,0.024)$ & $555.3 \pm 0.2$                      & $51.10 \pm 0.00$                    \\
        $D^- \to K_S^0 \pi^-$             & $(-0.025,0.026)$ & $65.6 \pm 0.1$                       & $51.42 \pm 0.01$                    \\
        $D^- \to K^+ \pi^- \pi^- \pi^0$   & $(-0.057,0.046)$ & $172.4 \pm 0.2$                      & $24.40 \pm 0.00$                    \\
        $D^- \to K_S^0 \pi^- \pi^0$       & $(-0.062,0.049)$ & $144.2 \pm 0.1$                      & $26.45 \pm 0.00$                    \\
        $D^- \to K_S^0 \pi^- \pi^- \pi^+$ & $(-0.028,0.027)$ & $79.0 \pm 0.1$                       & $29.59 \pm 0.01$                    \\
        $D^- \to K^+ K^- \pi^-$           & $(-0.024,0.023)$ & $48.1 \pm 0.1$                       & $40.91 \pm 0.01$                    \\
        \hline
        Sum                               &                  & $1064.7 \pm 0.4$                     &                                     \\
        \hline\hline
    \end{tabular}
    \label{tab:STyield}
\end{table}

\section{Event Selection and MC simulation}
Charged tracks detected in the MDC are required to be within a polar angle $(\theta)$ range of $|\cos \theta|<0.93$, where $\theta$ is defined with respect to the symmetry axis of the MDC ($z$) axis. For charged tracks not originating from $K_S^0$ decays, the distance of closest approach to the interaction point (IP) must be less than 10\,cm along the $z$ axis, $|V_z|$, and less than 1 cm in the transverse plane, $|V_{xy}|$. Particle identification (PID) for charged tracks combines measurements of the energy deposited in the MDC ($\text{d}E/\text{d}x$) and the flight time in the TOF to form likelihoods $\mathcal{L}(h)(h=p,K,\pi)$ for each hadron $h$ hypothesis. Charged kaons and pions are identified by comparing the likelihoods for the kaon and pion hypotheses, $\mathcal{L}(K)>\mathcal{L}(\pi)$ and $\mathcal{L}(\pi)>\mathcal{L}(K)$, respectively.

The $K_S^0\to\pi^+\pi^-$ candidate is reconstructed from two oppositely charged tracks satisfying $|V_z|<20~\text{cm}$. No PID requirement imposed on it. They are constrained to a common vertex, requiring an invariant mass within $(0.487,0.511)$~GeV/$c^2$. The decay length is required to be separated from the IP by more than twice resolution, which encompasses both the primary and secondary vertices. The quality of the vertex fits (primary vertex fit or secondary vertex fit) is ensured by a requirement on the $\chi^2(\chi^2<100)$.

The $\pi^0\to\gamma\gamma$ candidates are formed from EMC showers with deposited energy more than 25~MeV (barrel, $|\cos\theta|<0.80$) or larger than 50~MeV (endcap, $0.86<|\cos\theta|<0.92$), timing difference between the EMC time and the event start time within [0, 700]\,ns, angular separation $>10^\circ$ from any charged track, invariant mass within $[0.115,0.150],$GeV/$c^2$, and a kinematic fit is imposed to constrain its invariant mass to the known $\pi^0$ mass~\cite{pdg}. The $\chi^2$ of this kinematic fit is required to be less than 50.

Monte Carlo (MC) simulated data samples produced with a {\sc geant4}-based~\cite{geant4} software package, which includes the geometric description of the BESIII detector and the detector response, are used to determine detection efficiencies and to estimate backgrounds. The simulation models the beam energy spread and initial state radiation (ISR) in the $e^+e^-$ annihilations with the generator {\sc kkmc}~\cite{kkmc}. The inclusive MC sample includes the production of $D\bar{D}$ pairs (including quantum coherence for the neutral $D$ channels), the non-$D\bar{D}$ decays of the $\psi(3770)$, the ISR production of the $J/\psi$ and $\psi(3686)$ states, and the continuum processes incorporated in {\sc kkmc}~\cite{kkmc}.All particle decays are modelled with {\sc evtgen}~\cite{evtgen} using branching fractions (BFs) either taken from the Particle Data Group (PDG)~\cite{pdg}, when available, or otherwise estimated with {\sc lundcharm}~\cite{lundcharm}. Final state radiation~(FSR) from charged final state particles is incorporated using the {\sc photos} package~\cite{photos}.

\section{SYSTEMATIC uncertainties}

The differences between data and MC simulation are estimated with different control samples, and the obtained results are shown in Table~\ref{tab:corrfactor}. After correcting the signal efficiencies by these factors, their residual statistical uncertainties are assigned as the corresponding systematic uncertainties in the measurement of the branching fraction for $D^+\to K_S^0\pi^0\ell^+\nu_\ell$, as shown in Table~\ref{tab:sys}.

\begin{table}[htbp]\centering
    \caption{The used control samples and correction factors for different sources.}
    \label{tab:corrfactor}
    \begin{tabular}{cccc}
    \hline\hline
    Source                  & Control sample & $D^+ \to K^0_S\pi^0e^+\nu_e$ & $D^+ \to K^0_S\pi^0\mu^+\nu_\mu$ \\
    \hline 
    $\pi^+$ tracking        & $D^+\to K^-\pi^+\pi^+$, $D^0\to K^-\pi^+$ and $D^0\to K^-\pi^+\pi^+\pi^-$   & $0.994\pm0.002$& $0.994\pm0.002$\\
    $\pi^-$ tracking        & $D^+\to K^-\pi^+\pi^+$, $D^0\to K^-\pi^+$ and $D^0\to K^-\pi^+\pi^+\pi^-$   & $0.997\pm0.002$& $0.997\pm0.002$\\
    $K^0_S$ reconstruction  & $D^0\to K_S^0\pi^+\pi^-(\pi^0)$, $D^+\to K_S^0\pi^-\pi^0$ and $D^+\to K_S^0\pi^+\pi^-\pi^-$ & $0.991\pm0.002$& $0.992\pm0.002$\\
    $\pi^0$ reconstruction  & $D^0\to K^-\pi^+\pi^0$        & $0.976\pm0.002$& $0.977\pm0.002$\\
    $\ell^+$ tracking       & $e^+e^-\to \gamma e^+e^-$ and $e^+e^-\to \gamma \mu^+\mu^-$     & $1.005\pm0.002$ & $1.003\pm0.002$\\
    $\ell^+$ PID            & $e^+e^-\to \gamma e^+e^-$ and $e^+e^-\to \gamma \mu^+\mu^-$     & $0.974\pm0.002$ & $1.014\pm0.003$ \\
    $E_{\gamma}^\text{max},\,N_{\rm extra}^{\rm char},\,N_{\rm extra}^{\pi^0}$ 
                            & $D^+\to K^0_S\pi^0\pi^+\pi^0$ and $D^+\to K^0_S\pi^0\pi^+$  & $1.008\pm0.002$  & $1.018\pm0.002$  \\
    $M_{K_S^0\pi^0\mu}$ requirement & $D^+\to K^0_S\pi^0\pi^0\pi^+$ & $-$ & $0.999 \pm0.002$  \\
    \hline\hline
    \end{tabular}
\end{table}

\begin{table}[htbp]\centering
    \caption{The relative systematic uncertainties (in \%) in the branching fraction measurement.}
    \scalebox{1.0}{
    \begin{tabular}{lcc}
        \hline\hline
        Source                                     & $D^+\to K^0_S\pi^0 e^+\nu_e$ & $D^+\to K^0_S\pi^0 \mu^+\nu_\mu$ \\
        \hline
        $\ell^+$ tracking                          & 0.2         & 0.2                \\
        $\ell^+$ Particle ID                       & 0.2         & 0.3                \\
        $K_S^0$ reconstruction                     & 0.2         & 0.2                \\
        $\pi^0$ reconstruction                     & 0.2         & 0.2                \\
        $E_\gamma^{\max},\,N_{\rm extra}^{\rm track},\,N_{\rm extra}^{\pi^0}$
 requirements                                      & 0.6         & 0.6          \\
        $K_S^0\pi^0\ell^+$ mass window             & Negligible  & 0.2          \\
        $U_{\rm miss}$ fit                         & 0.7         & 0.3          \\
        $N^{\rm tot}_{\rm ST}$                     & 0.3         & 0.3          \\
        MC model                                   & 0.2         & 0.2          \\
        MC statistics                              & 0.2         & 0.2          \\
        Quoted branching fraction                  & 0.1         & 0.1          \\
        \hline
        Total                                      & 1.1         & 0.9          \\
        \hline
        \hline
    \end{tabular}
    }
    \label{tab:sys}
\end{table}

The systematic uncertainties of the hadronic transition form factors, $(K_S^0\pi^0)_{S-\text{wave}}$ fitted parameters and the fractions of $(K_S^0\pi^0)_{S-\text{wave}}$ and $\bar{K}^{*0}$ components are evaluated as the difference between the nominal fit results and those obtained after changing a variable or a condition by an amount corresponding to the estimated uncertainty of that quantity. The systematic uncertainties due to the background fraction is estimated by varying $f_{b}$ by $\pm 1\sigma$. The systematic uncertainties from the fixed parameters of $r_S^{(1)}$, $b^{1/2}_{\rm S,BG}$, and $r_{BW}$ are estimated by varying their input values by $\pm 1\sigma$~\cite{BESanfenfen}. The systematic uncertainties from the fixed parameters of $m_V$ and $m_A$ are estimated by varying their input values by $\pm 100$~MeV/$c^2$. To estimate the systematic uncertainty of efficiency corrections, we apply data-MC differences obtained from control samples for various components of the analysis (lepton tracking, lepton PID, $K^0_S$ reconstruction, $\pi^0$ reconstruction) to the MC weights. The effects of potential fit biases are evaluated by studying the pull distributions for the fitted parameters with a set of 300 toy MC samples. These differences are assigned as the systematic uncertainties, as summarized in Table~\ref{tab:Syserr}.

\begin{table}[htbp]\centering
    \caption{The relative systematic uncertainties in the amplitude analysis~(\%). For $\Delta m_{\bar{K}^{*0}}$, an additional energy calibration uncerainty is introduced as 0.2\%.}
    \resizebox{\linewidth}{!}{
    \begin{tabular}{ccccccccccccc}
    \hline\hline
    Source & $\Delta m_{\bar{K}^{*0}}$ & $\Delta \Gamma^0_{\bar{K}^{*0}}$ & $\Delta r_V$ & $\Delta r_2$ & $\Delta a(e)^{1/2}_{S,\text{BG}}$ & $\Delta a(\mu)^{1/2}_{S,\text{BG}}$ & $\Delta r_S(e)$& $\Delta r_S(\mu)$ & $\Delta f(e)_{S\text{-wave}}$ & $\Delta f(\mu)_{S\text{-wave}}$ & $\Delta f(e)_{\bar{K}^{*0}}$ & $\Delta f(\mu)_{\bar{K}^{*0}}$ \\\hline
    Background estimation   & 0.0 & 0.2 & 0.0 & 0.0 & 0.2 & 0.4 & 0.1 & 0.3 & 0.5 & 0.3 & 0.0 & 0.0\\
    $r_S^{(1)}$             & 0.0 & 0.5 & 0.5 & 0.1 & 0.6 & 0.2 & 3.8 & 5.6 & 1.8 & 4.3 & 0.1 & 0.4\\
    $b_{S,\text{BG}}^{1/2}$ & 0.0 & 0.1 & 0.4 & 1.1 & 10.6& 15.0& 0.2 & 0.3 & 0.7 & 0.0 & 0.0 & 0.0\\
    $r_{BW}$                & 0.0 & 0.7 & 0.6 & 0.1 & 0.8 & 0.3 & 2.0 & 0.6 & 0.9 & 3.8 & 0.1 & 0.3\\
    $m_{A}$                 & 0.0 & 0.0 & 0.6 & 1.1 & 0.1 & 0.0 & 0.1 & 0.4 & 0.1 & 0.4 & 0.0 & 0.0\\
    $m_{V}$                 & 0.0 & 0.0 & 1.1 & 0.0 & 0.0 & 0.1 & 0.0 & 0.0 & 0.3 & 0.4 & 0.0 & 0.0\\
    Efficiency corrections  & 0.0 & 0.0 & 0.1 & 0.1 & 0.1 & 0.1 & 0.1 & 0.1 & 0.1 & 0.1 & 0.1 & 0.1\\
    Fit bias                & 0.0 & 0.0 & 0.1 & 0.0 & 0.0 & 0.1 & 0.0 & 0.1 & 0.0 & 0.0 & 0.0 & 0.0\\ \hline
    Total                   & 0.2 & 1.0 & 1.6 & 1.7 & 10.6& 15.0& 4.4 & 6.1 & 2.2 & 5.8 & 0.3 & 0.5\\
    \hline\hline
    \end{tabular}
    \label{tab:Syserr}}
\end{table}

The sources of systematic uncertainties for the measurement of the angular coefficients are divided into three categories. The first one is from the yield determination, the second one is from efficiency determination and the last one is from the input parameters. For the yield determination, the uncertainty arises from the model used in the $U_\text{miss}$ fits, and is evaluated in the same manner as for the BF measurement. The systematic uncertainties associated with the efficiency determination are considered as two parts: one from the correction factor, which is estimated by varying the factor by $\pm 1\sigma$ (the effect is negligible), and another from the limited size of the signal MC sample.  Finally, the uncertainty from the input $D^-$ ST yield and $D^+$ lifetime are evaluated by varying their values by $\pm 1\sigma$, which are also negligible. The total uncertainties, obtained by adding the individual contributions in quadrature, are shown in Table~\ref{tab:asysys}. Note that the systematic uncertainties on the $A_{CP}$ results in Table~\ref{tab:asyAcp} are evaluated in a similar manner.  

\begin{table}[htbp]\centering
    \caption{The absolute systematic uncertainties for the angular observables $\left\langle A_i \right\rangle$ and $\left\langle S_i \right\rangle$ in $D^+ \to \bar{K}^{*0}\ell^+\nu_\ell$ across different $q^2$ regions with $0.8<m_{\bar{K}^{*0}}<1.0$ GeV$^2/c^4$.}
    \begin{tabular}{ccccccccc}
        \hline\hline
        $q^2$ (GeV$^2/c^4$) & $\langle A_2\rangle$ & $\langle A_3\rangle$ & $\langle A_4\rangle$ & $\langle A_5\rangle$ & $\langle A_6\rangle$ & $\langle A_7\rangle$ & $\langle A_8\rangle$ & $\langle A_9\rangle$ \\
        \hline
        \hline \multicolumn{9}{c}{$D^+ \to \bar{K}^{*0}e^+\nu_e$} \\\hline
        $[\,q^2_\text{min},0.20\,]$ &  0.1 & 0.4 & 0.4 & 0.3 & 0.4 & 0.2 & 0.2 & 0.2 \\
        $[\,0.20,0.40\,]$           &  0.1 & 0.1 & 0.1 & 0.2 & 0.1 & 0.2 & 0.3 & 0.1 \\
        $[\,0.40,0.60\,]$           &  0.1 & 0.2 & 0.1 & 0.2 & 0.2 & 0.1 & 0.1 & 0.1 \\
        $[\,0.60,0.80\,]$           &  0.3 & 0.4 & 0.3 & 0.2 & 0.3 & 0.0 & 0.6 & 0.1 \\
        $[\,0.80,q^2_\text{max}\,]$ &  0.5 & 0.3 & 0.4 & 1.1 & 0.1 & 0.1 & 0.6 & 0.3 \\
        Full range                  &  0.1 & 0.1 & 0.1 & 0.1 & 0.1 & 0.1 & 0.1 & 0.0 \\
        \hline
        \hline \multicolumn{9}{c}{$D^+ \to \bar{K}^{*0}\mu^+\nu_\mu$} \\\hline
        $[\,q^2_\text{min},0.20\,]$ &  1.3 & 0.3 & 2.1 & 0.3 & 0.6 & 0.7 & 0.3 & 0.4 \\
        $[\,0.20,0.40\,]$           &  0.4 & 0.2 & 0.2 & 0.5 & 0.2 & 0.7 & 0.2 & 0.9 \\
        $[\,0.40,0.60\,]$           &  0.5 & 0.3 & 0.4 & 0.1 & 0.1 & 0.2 & 0.2 & 0.2 \\
        $[\,0.60,0.80\,]$           &  0.3 & 0.5 & 1.2 & 0.2 & 0.8 & 0.3 & 0.7 & 0.5 \\
        $[\,0.80,q^2_\text{max}\,]$ &  0.4 & 3.9 & 1.2 & 4.3 & 0.0 & 0.4 & 0.0 & 1.7 \\
        Full range                  &  0.5 & 0.2 & 0.1 & 0.2 & 0.5 & 0.4 & 0.2 & 0.4 \\
        \hline\hline
\mbox{} \\
        \hline\hline
        $q^2$ (GeV$^2/c^4$) & $\langle S_2\rangle$ & $\langle S_3\rangle$ & $\langle S_4\rangle$ & $\langle S_5\rangle$ & $\langle S_6\rangle$ & $\langle S_7\rangle$ & $\langle S_8\rangle$ & $\langle S_9\rangle$ \\
        \hline
        \hline \multicolumn{9}{c}{$D^+ \to \bar{K}^{*0}e^+\nu_e$} \\\hline
        $[\,q^2_\text{min},0.20\,]$ &  1.1 & 0.2 & 0.6 & 1.3 & 2.9 & 0.1 & 0.2 & 0.4 \\
        $[\,0.20,0.40\,]$           &  0.8 & 0.2 & 0.6 & 0.2 & 0.7 & 0.2 & 0.5 & 0.2 \\
        $[\,0.40,0.60\,]$           &  0.7 & 0.2 & 0.3 & 0.4 & 0.4 & 0.1 & 0.0 & 0.3 \\
        $[\,0.60,0.80\,]$           &  0.4 & 0.2 & 0.5 & 0.2 & 0.2 & 0.1 & 0.3 & 0.0 \\
        $[\,0.80,q^2_\text{max}\,]$ &  1.0 & 0.4 & 0.4 & 1.1 & 0.1 & 0.3 & 0.6 & 0.1 \\
        Full range                  &  0.2 & 0.1 & 0.1 & 0.1 & 0.6 & 0.1 & 0.0 & 0.0 \\
        \hline
        \hline \multicolumn{9}{c}{$D^+ \to \bar{K}^{*0}\mu^+\nu_\mu$} \\\hline
        $[\,q^2_\text{min},0.20\,]$ &  1.0 & 1.1 & 1.7 & 0.2 & 1.3 & 0.5 & 0.7 & 0.6 \\
        $[\,0.20,0.40\,]$           &  0.2 & 0.3 & 0.5 & 0.7 & 0.8 & 0.1 & 0.2 & 0.2 \\
        $[\,0.40,0.60\,]$           &  0.1 & 0.2 & 0.9 & 0.3 & 0.6 & 0.4 & 0.2 & 0.1 \\
        $[\,0.60,0.80\,]$           &  0.3 & 0.2 & 2.2 & 0.3 & 0.3 & 0.0 & 0.1 & 0.4 \\
        $[\,0.80,q^2_\text{max}]$   &  0.4 & 4.3 & 0.3 & 4.6 & 0.8 & 0.4 & 0.3 & 1.9 \\
        Full range                  &  0.3 & 0.2 & 0.1 & 0.4 & 0.3 & 0.3 & 0.1 & 0.3 \\
        \hline\hline
    \end{tabular}
    \label{tab:asysys}
\end{table}

\end{document}

%% file: authorlist_2025-01-22.tex
\author{
\small
M.~Ablikim$^{1}$, M.~N.~Achasov$^{4,c}$, P.~Adlarson$^{77}$, X.~C.~Ai$^{82}$, R.~Aliberti$^{36}$, A.~Amoroso$^{76A,76C}$, Q.~An$^{73,59,a}$, Y.~Bai$^{58}$, O.~Bakina$^{37}$, Y.~Ban$^{47,h}$
% ...
\\
\vspace{0.2cm}
(BESIII Collaboration)\\
\vspace{0.2cm} {\it
$^{1}$ Institute of High Energy Physics, Beijing 100049, People's Republic of China\\
% ...
$^{4}$ Budker Institute of Nuclear Physics SB RAS (BINP), Novosibirsk 630090, Russia\\
% ...
$^{36}$ Johannes Gutenberg University of Mainz, Johann-Joachim-Becher-Weg 45, D-55099 Mainz, Germany\\
$^{37}$ Joint Institute for Nuclear Research, 141980 Dubna, Moscow region, Russia\\
% ...
$^{47}$ Peking University, Beijing 100871, People's Republic of China\\
% ...
$^{58}$ Southeast University, Nanjing 211100, People's Republic of China\\
$^{59}$ State Key Laboratory of Particle Detection and Electronics, Beijing 100049, Hefei 230026, People's Republic of China\\
% ...
$^{73}$ University of Science and Technology of China, Hefei 230026, People's Republic of China\\
% ...
$^{76}$ University of Turin and INFN, (A)University of Turin, I-10125, Turin, Italy; (B)University of Eastern Piedmont, I-15121, Alessandria, Italy; (C)INFN, I-10125, Turin, Italy\\
$^{77}$ Uppsala University, Box 516, SE-75120 Uppsala, Sweden\\
% ...
$^{82}$ Zhengzhou University, Zhengzhou 450001, People's Republic of China\\
\vspace{0.2cm}
$^{a}$ Deceased\\
% ...
$^{c}$ Also at the Novosibirsk State University, Novosibirsk, 630090, Russia\\
% ...
$^{h}$ Also at State Key Laboratory of Nuclear Physics and Technology, Peking University, Beijing 100871, People's Republic of China\\
% ...
}
}

%% file: acknowledgement_2025-01-22.tex
\textbf{Acknowledgement}

The BESIII Collaboration thanks the staff of BEPCII (https://cstr.cn/31109.02.BEPC) and the IHEP computing center for their strong support. This work is supported in part by National Key R\&D Program of China under Contracts Nos. 2023YFA1606000, 2023YFA1606704; National Natural Science Foundation of China (NSFC) under Contracts Nos. 11635010, 11935015, 11935016, 11935018, 12025502, 12035009, 12035013, 12061131003, 12192260, 12192261, 12192262, 12192263, 12192264, 12192265, 12221005, 12225509, 12235017, 12361141819; the Chinese Academy of Sciences (CAS) Large-Scale Scientific Facility Program; CAS under Contract No. YSBR-101; 100 Talents Program of CAS; The Institute of Nuclear and Particle Physics (INPAC) and Shanghai Key Laboratory for Particle Physics and Cosmology; Agencia Nacional de Investigación y Desarrollo de Chile (ANID), Chile under Contract No. ANID PIA/APOYO AFB230003; German Research Foundation DFG under Contract No. FOR5327; Istituto Nazionale di Fisica Nucleare, Italy; Knut and Alice Wallenberg Foundation under Contracts Nos. 2021.0174, 2021.0299; Ministry of Development of Turkey under Contract No. DPT2006K-120470; National Research Foundation of Korea under Contract No. NRF-2022R1A2C1092335; National Science and Technology fund of Mongolia; National Science Research and Innovation Fund (NSRF) via the Program Management Unit for Human Resources \& Institutional Development, Research and Innovation of Thailand under Contract No. B50G670107; Polish National Science Centre under Contract No. 2024/53/B/ST2/00975; Swedish Research Council under Contract No. 2019.04595; U. S. Department of Energy under Contract No. DE-FG02-05ER41374

%% file: dkstarlnu.bbl
\begin{thebibliography}{99}
\bibitem{illustrate}
    Throughout this Letter, charged conjugations are always implied and the notation of $\bar{K}^{*0}$ indicates the $\bar{K}^*(892)^0$.
\end{thebibliography}
